\documentclass[useAMS,usenatbib]{mn2e}

\usepackage{longtable,natbib,amssymb}
\usepackage{graphicx}
\usepackage{times}

\def\arcsec{{\mbox{$^{\prime \prime}$}}}

\def\degree{{\mbox{$^{\circ}$}}}

\def \deg     {\ifmmode^\circ\else$^\circ$\fi}  
\def \solar   {\ifmmode_{\mathord\odot} \else$_{\mathord\odot}$\fi}
\def \Msun   {{\rm M}\solar}

\def\arcsec{{\mbox{$^{\prime \prime}$}}}

\def\erg{{\rm\thinspace erg}}

\def\km{{\rm\thinspace km}}

\def\100{\hbox{$\rm\thinspace M_{\odot}$}}

\def\s{{\rm\thinspace s}}

\def\ergps{\hbox{$\erg\s^{-1}\,$}}

\def\kmps{\hbox{$\km\s^{-1}\,$}}

\def\arcsec{{\mbox{$^{\prime \prime}$}}}

\def\degree{{\mbox{$^{\circ}$}}}

\title[SDSS1133+5504: An Unusually Persistent Transient in a Nearby Dwarf Galaxy]{SDSS1133: An Unusually Persistent Transient in a Nearby Dwarf Galaxy}
\author[M.~Koss et al.]
{\parbox{\textwidth}{Michael Koss$^{1,2}$\thanks{E-mail: \texttt{mkoss@phys.ethz.ch}},
Laura Blecha$^{3,4}$,
Richard Mushotzky$^{3}$,
Chao Ling Hung$^{2}$,
Sylvain Veilleux$^{3}$,
Benny Trakhtenbrot$^{1,5}$,
Kevin Schawinski$^{1}$,
Daniel Stern$^{6}$,
Nathan Smith$^{7}$,
Yanxia Li$^{2}$,
Allison Man$^{8}$,
Alexei V. Filippenko$^{9}$,
Jon C. Mauerhan$^{7}$,
Kris Stanek$^{10}$, and
David Sanders$^{2}$}\vspace{0.4cm}\\
\parbox{\textwidth}{$^{1}$Institute for Astronomy, Department of Physics, ETH Zurich, Wolfgang-Pauli-Strasse 27, CH-8093 Zurich, Switzerland\\
$^{2}$Institute for Astronomy, University of Hawaii, 2680 Woodlawn Drive, Honolulu, HI 96822, USA\\
$^{3}$Astronomy Department, University of Maryland, College Park, MD, USA\\
$^{4}$Einstein Fellow\\
$^{5}$Zwicky Fellow\\
$^{6}$Jet Propulsion Laboratory, California Institute of Technology, 4800 Oak Grove Drive, MS 169-506, Pasadena, CA 91109, USA\\
$^{7}$Steward Observatory, University of Arizona, 933 N. Cherry Ave., Tucson, AZ 85721, USA\\
$^{8}$Dark Cosmology Center, University of Copenhagen, Copenhagen, Denmark\\
$^{9}$Department of Astronomy, University of California, Berkeley, CA 94720-3411, USA\\
$^{10}$Department of Astronomy, The Ohio State University, Columbus, Ohio 43210, USA}}

\begin{document}

\pagerange{\pageref{firstpage}--\pageref{lastpage}} 
\maketitle
\label{firstpage}

\begin{abstract}
While performing a survey to detect recoiling supermassive black holes, we have identified an unusual source having a projected offset of 800 pc from a nearby dwarf galaxy.  The object,\\ SDSS J113323.97+550415.8, exhibits broad emission lines and strong variability. While originally classified as a supernova (SN) because of its nondetection in 2005, we detect it in recent and past observations over 63 yr and find over a magnitude of rebrightening in the last 2 years.  Using high-resolution adaptive optics observations, we constrain the source emission region to be $\lesssim 12$ pc and find a disturbed host-galaxy morphology indicative of recent merger activity.  Observations taken over more than a decade show narrow [O~III] lines, constant ultraviolet emission, broad Balmer lines, a constant putative black hole mass over a decade of observations despite changes in the continuum, and optical emission-line diagnostics consistent with an active galactic nucleus (AGN).  However, the optical spectra exhibit blueshifted absorption, and eventually narrow Fe~II and [Ca~II] emission, each of which is rarely found in AGN spectra.  While this peculiar source displays many of the observational properties expected of a potential black hole recoil candidate, some of the properties could also be explained by a luminous blue variable star (LBV) erupting for decades since 1950, followed by a Type IIn SN in 2001.  Interpreted as an LBV followed by a SN analogous to SN 2009ip, the multi-decade LBV eruptions would be the longest ever observed, and the broad H$\alpha$ emission would be the most luminous ever observed at late times ($>10$ yr), larger than that of unusually luminous supernovae such as SN 1988Z, suggesting one of the most extreme episodes of pre-SN mass loss ever discovered.
\end{abstract}


\section{Introduction}

       The coalescence of binary supermassive black holes (SMBHs) in galaxy mergers is thought to constitute the strongest source of gravitational waves \citep[GW;][]{Merritt:2005:8}.  Theory suggests that these waves carry momentum, causing the merged black hole (BH) to experience a velocity recoil or kick that displaces it from the center of its host galaxy \citep{Peres:1962:2471,Bekenstein:1973:657,Thorne:1976:L1}.  Numerical simulations of SMBH mergers have found, surprisingly, that GW recoil kicks may be quite large, up to $\sim 5000~\kmps$ \citep{Campanelli:2006:11207, Lousto:2011:231102}. Consequently, a merged SMBH may even be ejected from its host galaxy. 

        An active galactic nucleus (AGN) ejected from the center of a galaxy should be able to carry along its accretion disk as well as the broad-line region, resulting in an AGN spatially offset from its host galaxy and/or an AGN with broad emission lines offset in velocity \citep{Madau:2004:L17, Komossa:2008:L89, Blecha:2008:1311}. Recoiling AGNs with offsets $> 1$ kpc may have lifetimes up to tens of Myr for a fairly wide range in kick speeds, and velocity-offset AGNs may have similar lifetimes \citep{Blecha:2011:2154,Blecha:2013:1341}. Thus far, several candidate recoiling SMBHs have been found via spatial offsets, though none has been confirmed \citep[see][for a review]{Komossa:2012:14}. The disturbed galaxy CXOC J100043.1+020637 (or CID-42) contains a candidate recoiling AGN offset by 2.5 kpc from the galactic center \citep{Civano:2010:209,Civano:2012:49}. However, in order for the recoiling AGN to produce narrow-line emission, it must be observed very quickly after the kick while it still inhabits a dense gaseous region.  Another recoil candidate found by \citet{Jonker:2010:645} consists of an X-ray source offset by 3 kpc from the center of an apparently undisturbed spiral galaxy, but this could also be explained as an ultraluminous X-ray source (ULX) associated with an accreting intermediate-mass black hole (IMBH) in a massive, young stellar cluster, or perhaps a very luminous Type IIn supernova \citep[SN~IIn; see][for a review]{Filippenko:1997:309}.
	
	  These recoiling kicks have significant implications for models of SMBH and galaxy coevolution \citep{Volonteri:2007:L5,Sijacki:2011:3656,Blecha:2011:2154}.  A hitherto unknown low-redshift population of SMBH mergers in dwarf galaxies could have profound implications for gravitational wave detectors.  Although the SMBH occupation fraction in dwarf galaxies is uncertain, their shallower potential wells greatly increase the possibility of detecting gravitational wave kicks; nearly every major SMBH merger would result in a substantial displacement of the remnant SMBH.  While current detectors, like the advanced {\it Laser Interferometer Gravitational-Wave Observatory} or Pulsar Timing Arrays, are more sensitive to stellar-mass BHs or the most massive SMBHs \citep{Sesana:2013:4086}, future space-based detectors such as the proposed Laser Interferometer Space Antenna would be more sensitive to lower-mass SMBHs ($\sim10^5$--$10^6~\Msun$), making this mass range found in dwarf galaxies an important population for study.

	Particularly at low bolometric luminosities ($L_{\rm bol} \le 10^{43}$ $\ergps$), some Type II supernovae (SNe) can be a source of imposters for AGNs \citep{Filippenko:1989:726} and recoiling SMBHs, especially when the broad-line emission lasts for decades. Some SNe are also preceded by strong winds and nonterminal eruptions similar to those of luminous blue variables (LBVs) such as $\eta$ Carinae \citep{Smith:2011:773}.  There is also evidence that LBV eruptions can be followed by SNe~IIn \citep{Gal-Yam:2007:372, Mauerhan:2013:1801,Ofek:2014}, with narrow as well as broader Balmer emission lines that can persist for decades \citep{Smith:2008:467,Kiewe:2012:10}, sometimes with an appearance similar to that of broad-line AGNs. SNe~IIn occur in a dense circumstellar medium (CSM) created by pre-SN mass loss, and the narrow hydrogen lines are produced by photoionization of the dense winds irradiated by X-rays from the region behind the forward shock \citep{Chevalier:1994:268}. (The broader lines arise when the SN ejecta interact directly with the CSM.)  Type IIn SNe can even produce forests of narrow coronal lines such as [O~III] \citep[e.g., SN 2005ip;][]{Smith:2009:1334}. 

	Our study focuses on an enigmatic point source, SDSS J113323.97+550415.8 (hereafter SDSS1133), having a projected offset of 800 pc (5.8$\arcsec$) from the center of a nearby dwarf galaxy, Mrk 177 (UGC 239), and with broad-line emission observed by the Sloan Digital Sky Survey (SDSS) in 2003.  SDSS1133 was mentioned as a possible case of a quasar having a noncosmological redshift because of its broad Balmer lines but very low luminosity \citep{Lopez-Corredoira:2005}.  In a study of narrow-line Seyfert 1 galaxies, \citet{Zhou:2006:128} classified SDSS1133 as a SN because of its nondetection in data obtained in January 2005 with the 2.16~m telescope at the Beijing Observatory.  This SN classification was also applied by \citet{Reines:2013:116} based on the SDSS spectra using the automated SN detection code of \citet{Graur:2013:1746}.  Finally, SDSS1133 was listed as a possible ``Voorwerp" candidate for a giant ionized cloud \citep{Keel:2012:878}. 

        To better understand this unusual source, we obtained new optical spectra, adaptive optics (AO) images, and ultraviolet (UV) and X-ray observations of it. We also analyzed archival images, finding that the object is detected over a time span of 63 yr. The host galaxy, Mrk 177, is at a distance of 28.9 Mpc \citep[distance modulus 32.3 mag;][]{Tully:1994}.  We use this redshift-independent distance indicator for the subsequent analysis of SDSS1133 and Mrk 177.  At this distance, 1$\arcsec$ corresponds to 140 pc.  Galactic foreground extinction is very low, $A_V = 0.03$ mag \citep{Schlafly:2011:103}.

\section{Observations and Data Analysis}

\subsection{Imaging}

	We use archival optical SDSS images from 2001 December 18 and 2002 April 1 (UT dates are used throughout this paper), as well as 111 $griz$ images taken over 26 nights from Pan-STARRS1 (PS1) between 2010 March and 2014 March.  Additionally, we have photographic Digital Sky Survey (DSS) plates from the POSS I and II surveys, with 1$\arcsec$ pixel$^{-1}$ sampling and a limiting magnitude of 21 and 22.5, respectively.  The 1994 observation, with the IIIaJ emulsion and GG395 filter, is comparable to the $g$ filter.  The 1950 plate from the POSS-I O survey uses the 103aO emulsion with a response between $u$ and $g$.  The 1999 observation is with the IIIaF emulsion, similar to the $i$ band.  Example images can be found in Figure~\ref{fig:HistoricalImage}.
	
	We measure the photometry of SDSS1133 (Fig.~2--3) using a two-dimensional surface brightness fitting program, GALFIT \citep{Peng:2002:266}.  We fit a 5$\arcsec$ region around SDSS1133, using a point-spread function (PSF) model for the source light and a linear model based on radial distance to model contamination from Mrk 177.  The PSF model position is allowed to vary across the image region.  PSF models from the SDSS and PS1 data are used for the image convolution.  For the PSF model in DSS images, we use a median of 5 nearby, bright, unsaturated stars.  The DSS imaging scale is in photographic density, which is nonlinear with intensity.  Thus, the increase in contamination from the galaxy causes the source intensity to be underestimated.  To correct for this nonlinearity in DSS images, we measure SDSS PSF magnitudes of 5 nearby unsaturated stars with varying brightnesses to calculate the magnitude of SDSS1133 in the DSS plates.
	
	We fit the positions of 5 bright, unsaturated stars common to each of the images for source astrometry, using the 2001 SDSS astrometry as the reference.  We find no significant difference between SDSS and PS1 astrometry.  For the DSS images, the declination offsets are $<0.2\arcsec$ with respect to the SDSS astrometry, while we find right-ascension offsets of 1$\arcsec$ in the 1950 DSS image and 0.6$\arcsec$ in the 1994 image.

	We also estimate photometry from SDSS spectra taken in March 2003 using PYSYNPHOT, which determines the flux in the spectra as measured by different photometric filters.  We convert from spectra filter magnitudes to PSF magnitudes using the offset between fibermag and PSF magnitudes from the original SDSS images.  Finally, we include as an upper limit an observation in January 2005 using the 2.16-m telescope of the Beijing Observatory which failed to detect the source within 3 mag of the 2002 SDSS observation \citep{Zhou:2006:128}.  

	We obtained AO images on 2013 June 16 with the NIRC2 instrument on the Keck-2 10-m telescope equipped with the wide camera, which has a 40$\arcsec$ field of view (FOV) and 40 mas pixel$^{-1}$ (Fig.~4).   In this observation, we used a 3-point dither pattern for 6 min in the $J$ and $K_p$ filters, as well as 12 min in the Pa$\beta$ filter.  The Pa$\beta$ filter covers the redshift of SDSS1133 and Mrk 177 ($z=0.007845$).  An image of the AO tip-tilt star is used for PSF estimation (Fig.~5) and photometric calibration based on measured 2MASS magnitudes in $J$ and $K_p$.
	
	The {\it Swift} satellite observed SDSS1133 as a Target of Opportunity (ToO) program on 2013 August 26, 27, and 28 for 3.5, 3.8, and 11.2 ks, respectively, with its X-Ray Telescope (XRT) in Photon Counting mode (PC mode) and the UV-Optical Telescope (UVOT) with the UVW1 near-UV filter.  The total exposure time was 18.5 ks.  X-ray data were reduced with the task XRTPIPELINE (version 0.12.6). Source and background photons were extracted with XSELECT (version 2.4b), from circles with radii of 47$\arcsec$ and 200$\arcsec$, respectively, in the 0.3--10 keV band.  For PSF photometry with UVOT, we use a nearby star with known magnitudes scaled to the brightness of SDSS1133.

\subsection{Optical Spectroscopy}	

	We use optical spectra from a variety of telescopes for this study of SDSS1133 (Fig.~6--8).  SDSS spectra were taken in 2003 March, 456 days after the object had diminished in brightness about 2.5 mag from its peak in 2001 December.  SDSS1133 was targeted as part of the $ugri$-selected quasar survey because of its AGN-like colors.  
	
        SDSS1133 was also observed with the University of Hawaii 2.2~m telescope and the SuperNova Integral Field Spectrograph (SNIFS) on 1--4 May 2013 for a total duration of 160~min.  SNIFS is an optical integral field unit (IFU) spectrograph with blue (3000--5200~\AA) and red (5200--9500~\AA) channels having a resolution of 360 $\kmps$.  The SNIFS reduction pipeline SNURP was used for wavelength calibration, spectro-spatial flatfielding, cosmic ray removal, and flux calibration \citep{Aldering:2006:510}.  A sky image was taken after each source image and subtracted from each IFU observation.  Flux corrections were applied each night based on the standard star Feige 34. In order to estimate the amount of narrow-line contamination from Mrk 177 at the position of SDSS1133 in the SDSS spectrum, we placed an aperture at the same radial distance from Mrk 177, but to the northeast of SDSS1133 with the same aperture size as the SDSS fiber (3$\arcsec$).

         We used the DEIMOS spectrograph on the Keck-II telescope to obtain a 10~min spectrum (range 4730--9840~\AA) of the object and the galaxy nucleus (2$\arcsec$ slit, PA = 137.5$\degree$) with a 600 lines mm$^{-1}$ grating on 2013 December 13. Finally, we used the MMT to observe SDSS1133 on 2014 January 3 (1$\arcsec$ slit, PA = 137.5$\degree$); the spectrum spans 3700--8926~\AA.

	We fit the spectra using an extensible spectroscopic analysis toolkit for astronomy, PYSPECKIT, which uses a Levenberg-Marquardt algorithm for fitting.  We adopt a power-law fit to model the continuum and Gaussian components to model the emission lines.  All narrow-line widths were fixed to the best fit of the [O~III] $\lambda$5007 line.  We fit the spectra of H$\alpha$ using a narrow component based on the [O~III] $\lambda$5007 line along with a broad component.  Additionally, some AGNs have broad Balmer lines that are poorly fit with a single Gaussian component, so we fit H$\alpha$ and H$\beta$ with two broad components (broad and very broad) when it is statistically significant based on the reduced $\chi^2$.  Fitting the H$\beta$ line is more complicated because of Fe~II lines on the red wing as well as [O~III] $\lambda$4959 on the red wing; we therefore use the H$\beta$ fitting procedure following the code described by \citet{Trakhtenbrot:2012:3081}.    To measure the full width at half-maximum intensity (FWHM) of the broad component of the Balmer lines, we first subtract the narrow-line component.  Finally, we fit a separate broad component to each of the calcium near-infrared (NIR) triplet ($\lambda\lambda$8498, 8452, 8662) lines and to O~I $\lambda$8446.  The luminosities of the narrow lines and their velocity offsets from Mrk 177 can be found in Table 2.

\section{Results}

\subsection{X-ray, UV, Optical, and NIR Emission}

\begin{figure*}
\includegraphics[width=4.2cm]{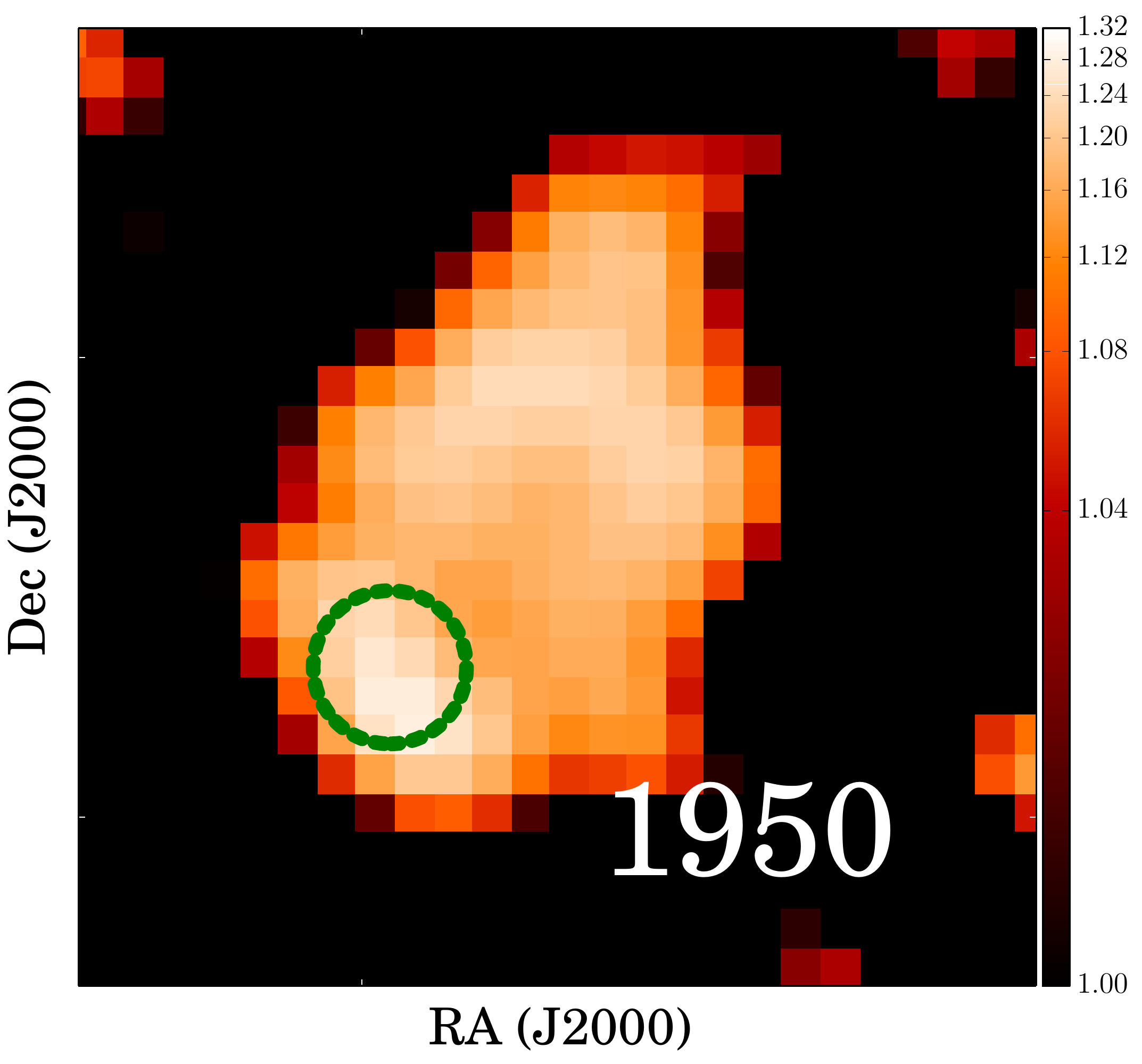}
\includegraphics[width=4.2cm]{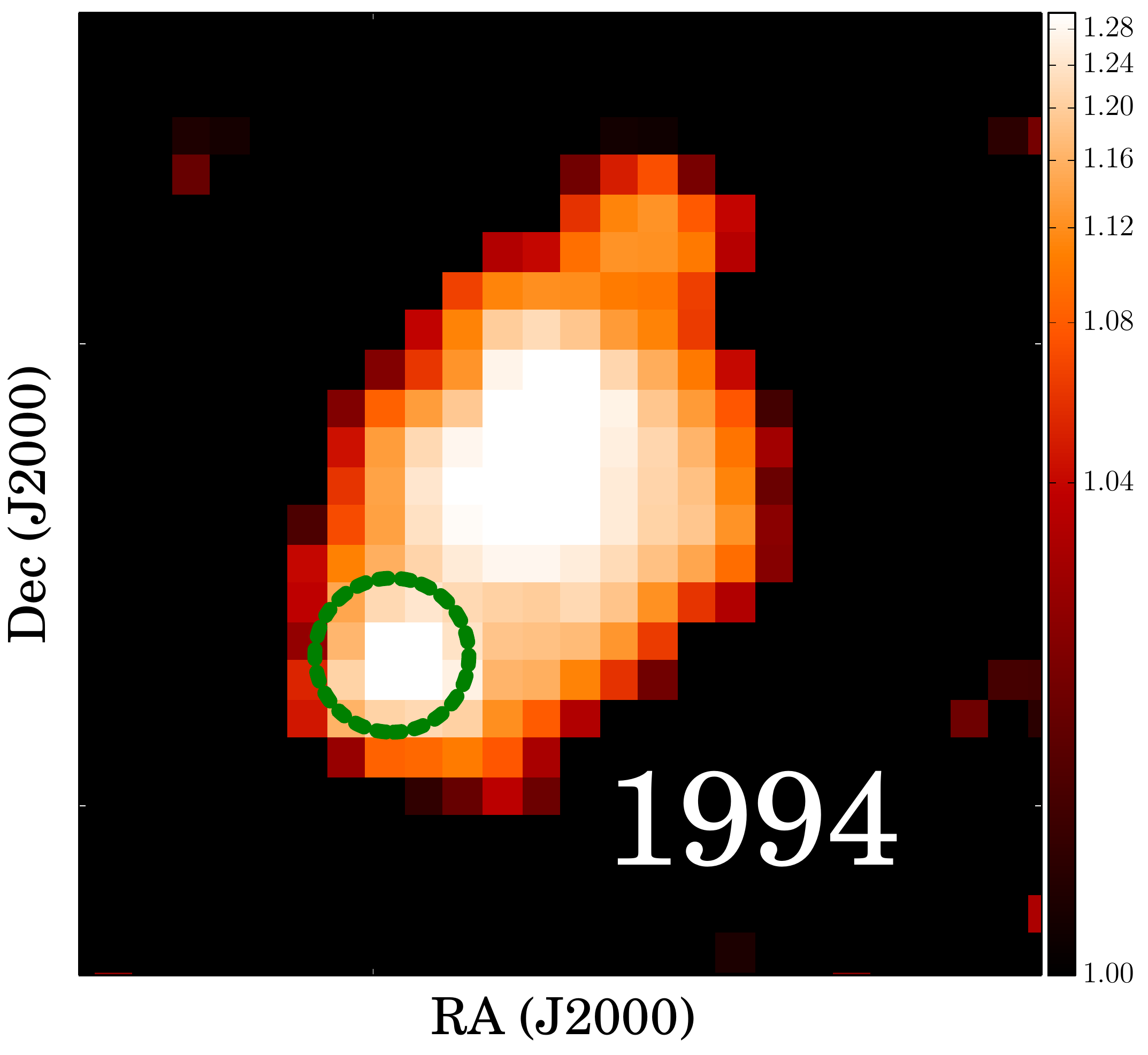}
\includegraphics[width=4.2cm]{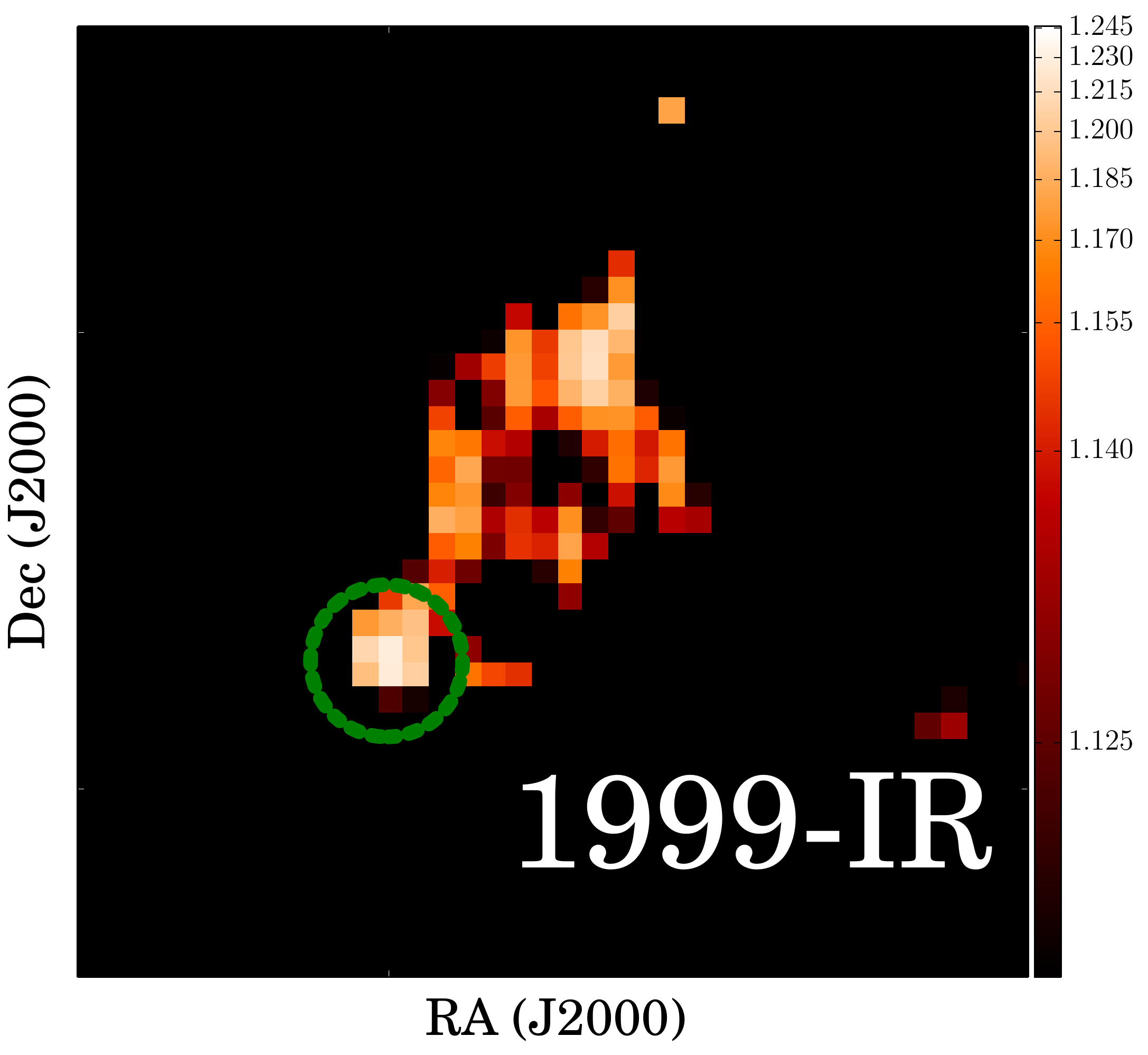}
\includegraphics[width=4.2cm]{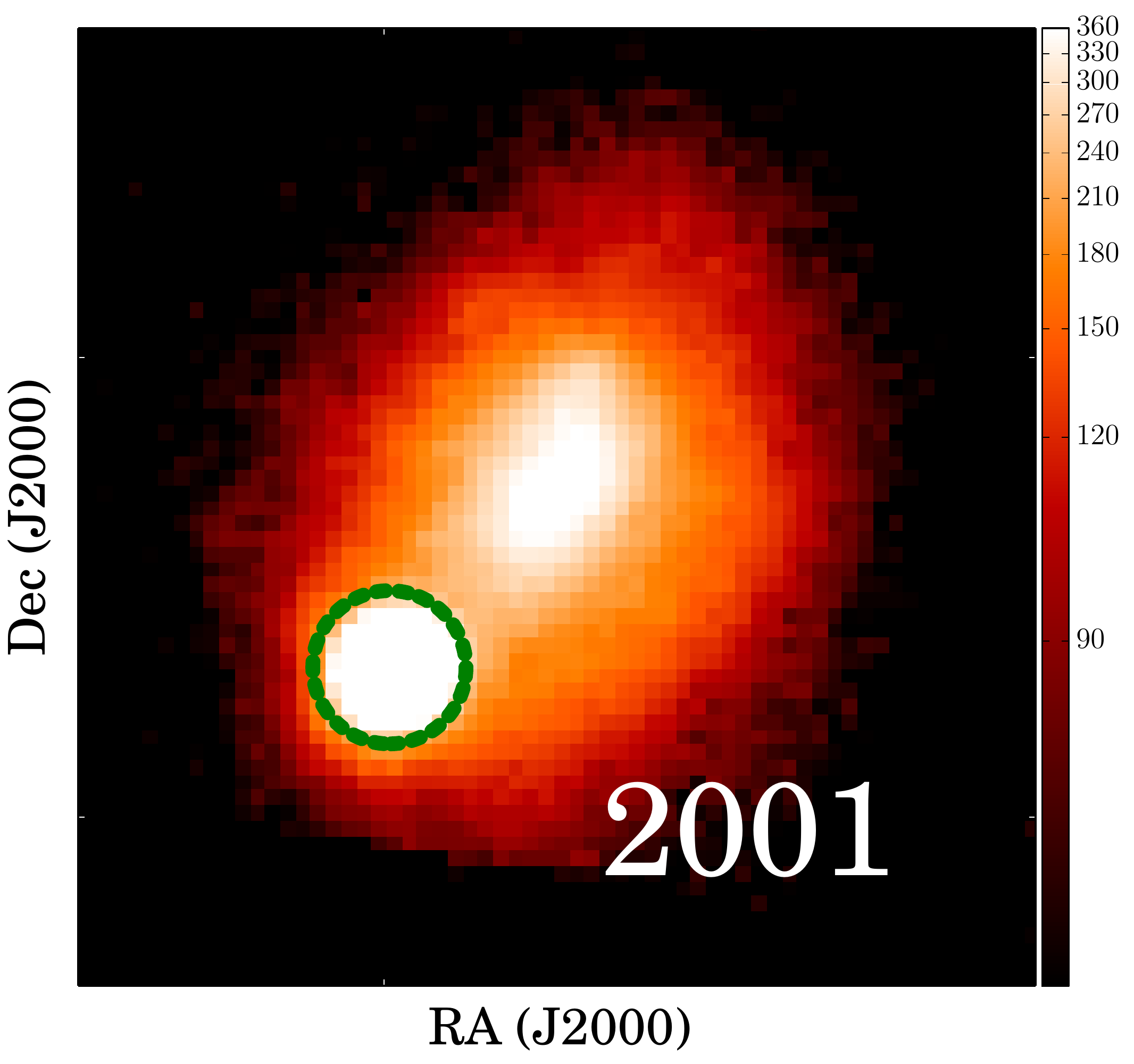}
\includegraphics[width=4.2cm]{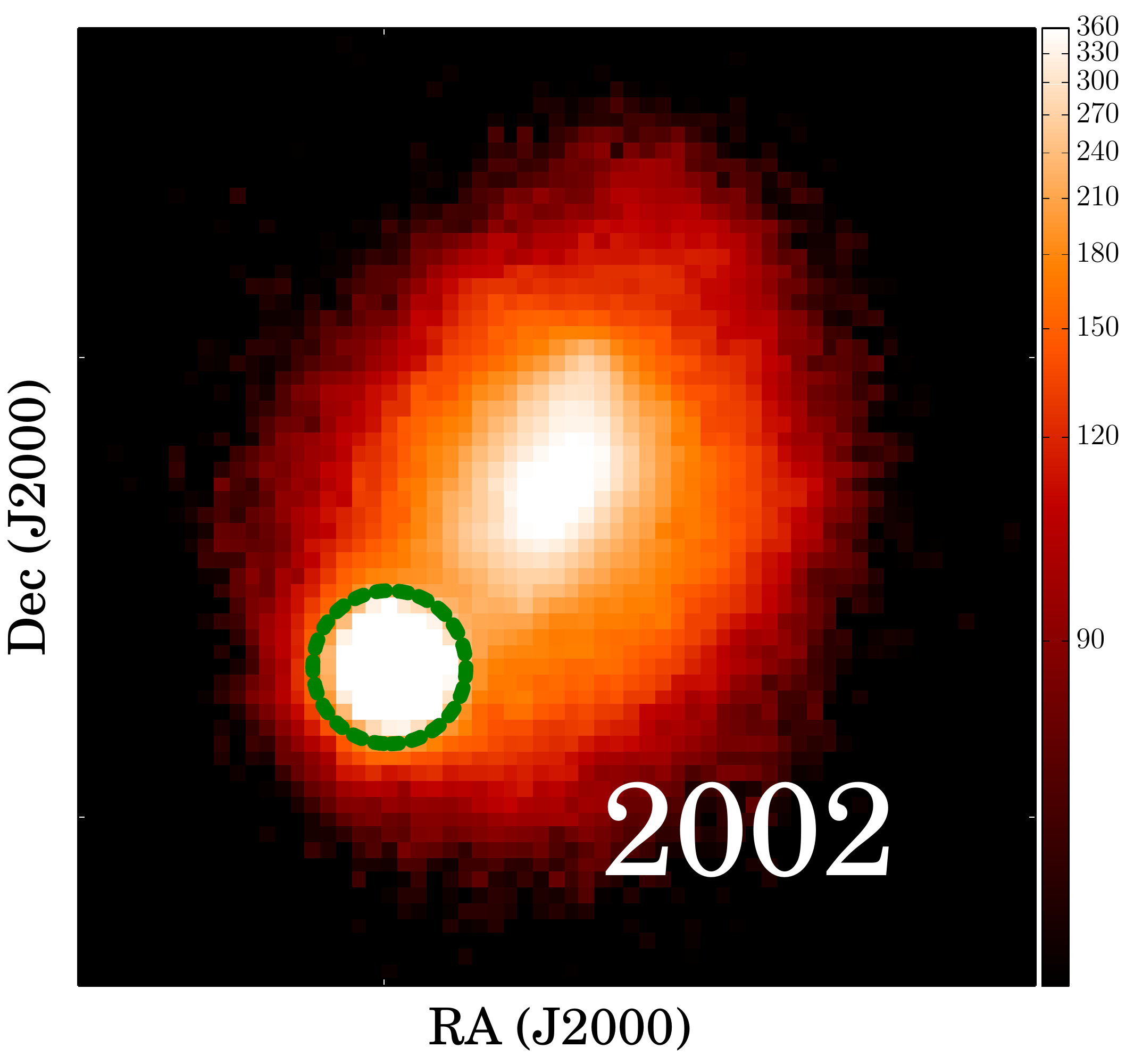}
\includegraphics[width=4.2cm]{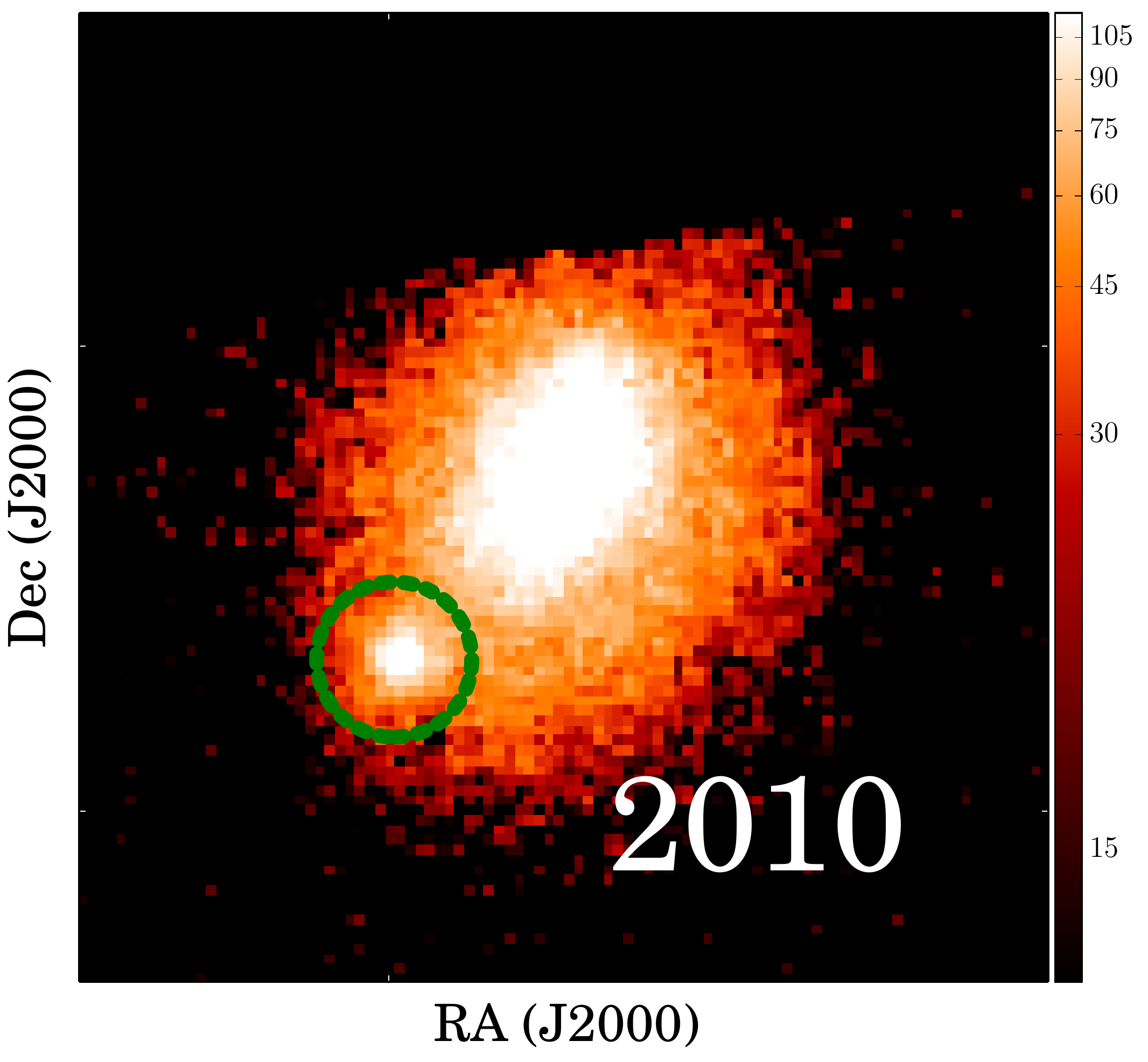}
\includegraphics[width=4.2cm]{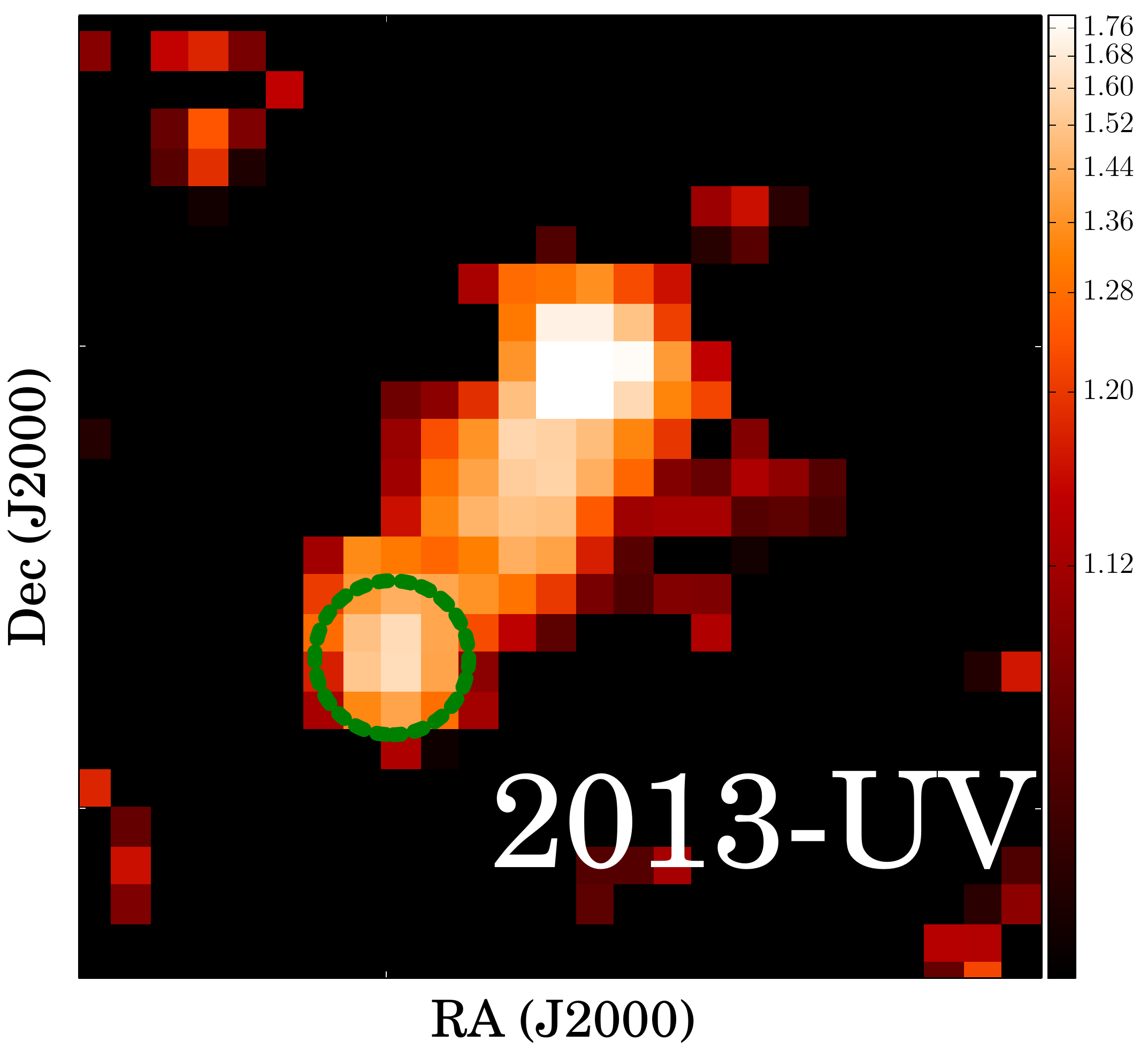}
\includegraphics[width=4.2cm]{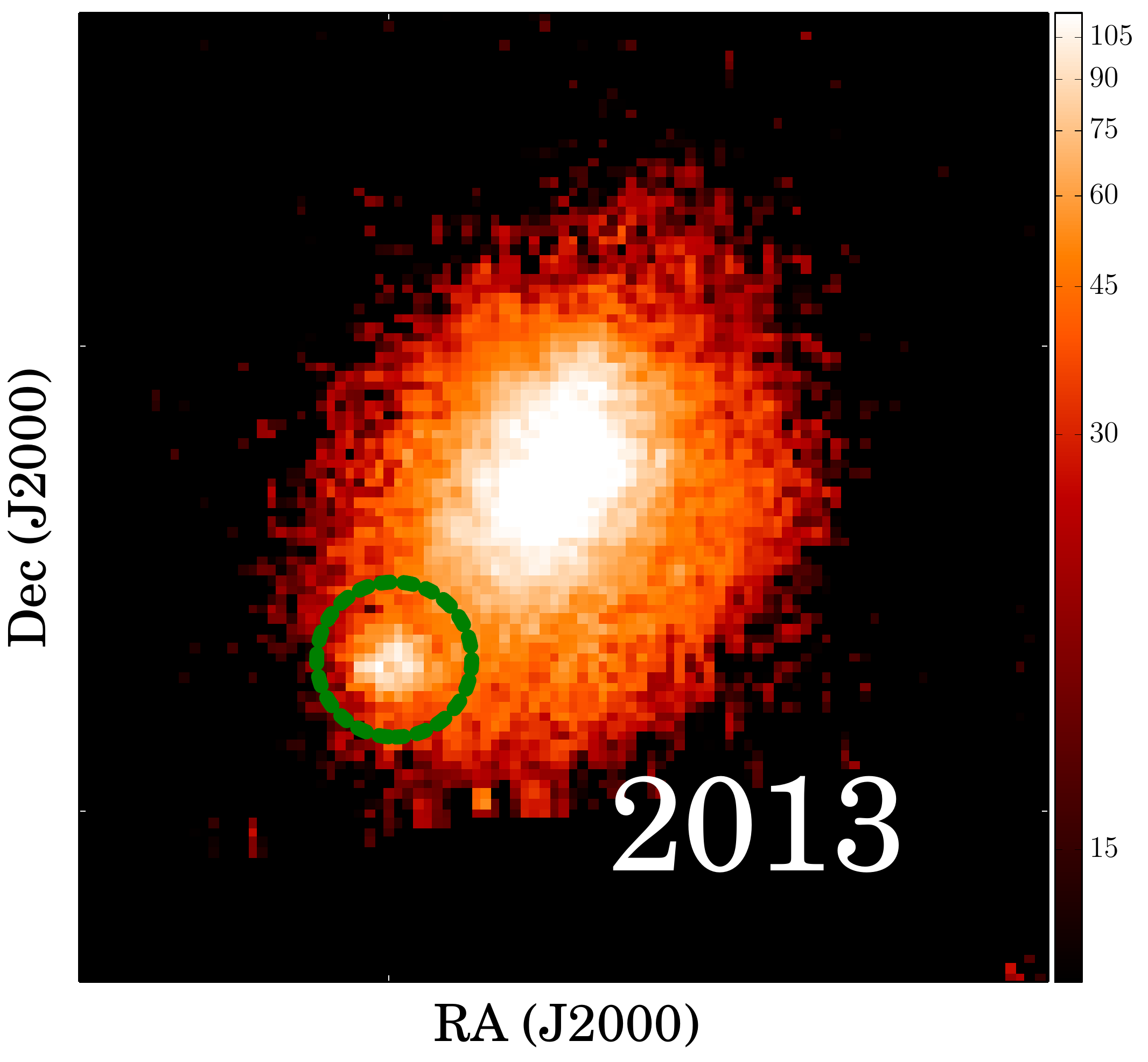}
\caption{Images of the host galaxy Mrk 177 and SDSS1133, 25$\arcsec$ wide and displayed with an arcsinh scale representing photographic density or CCD counts.  The 1950 and 1994 images are from blue DSS plates, the 1999 image is from an IR DSS plate, and the 2001--2013 data are $g$-band images from the SDSS and PS1.  A dashed green circle of 2$\arcsec$ radius is drawn for SDSS1133 based on positions in the 2001 SDSS observation.  A 3-pixel unsharp mask has been applied to the DSS and UV images because of their lower resolution and pixel sampling.}
\label{fig:HistoricalImage}
\end{figure*}

\begin{figure*}
\centering
\includegraphics[width=8.1cm]{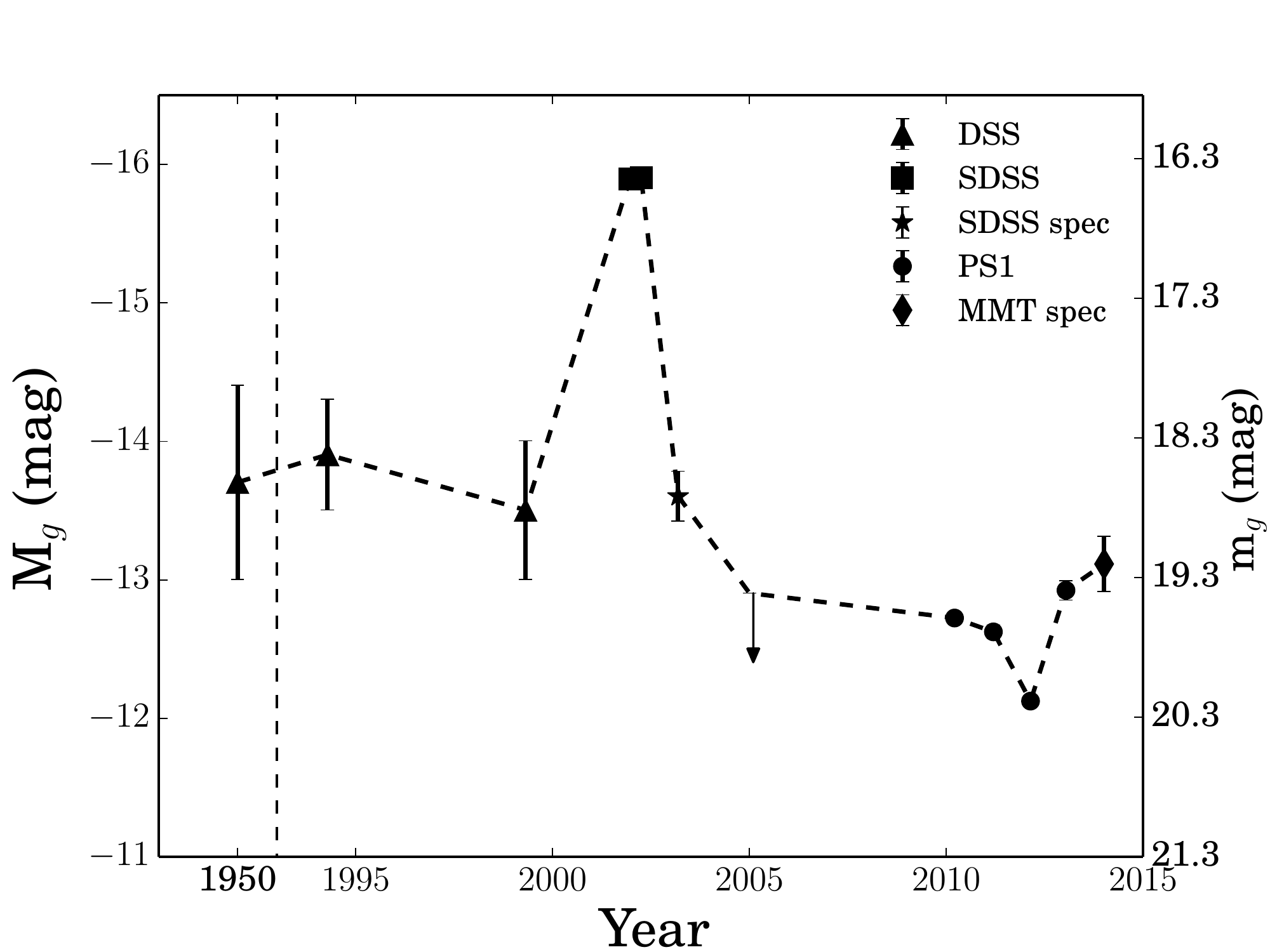}
\includegraphics[width=8.1cm]{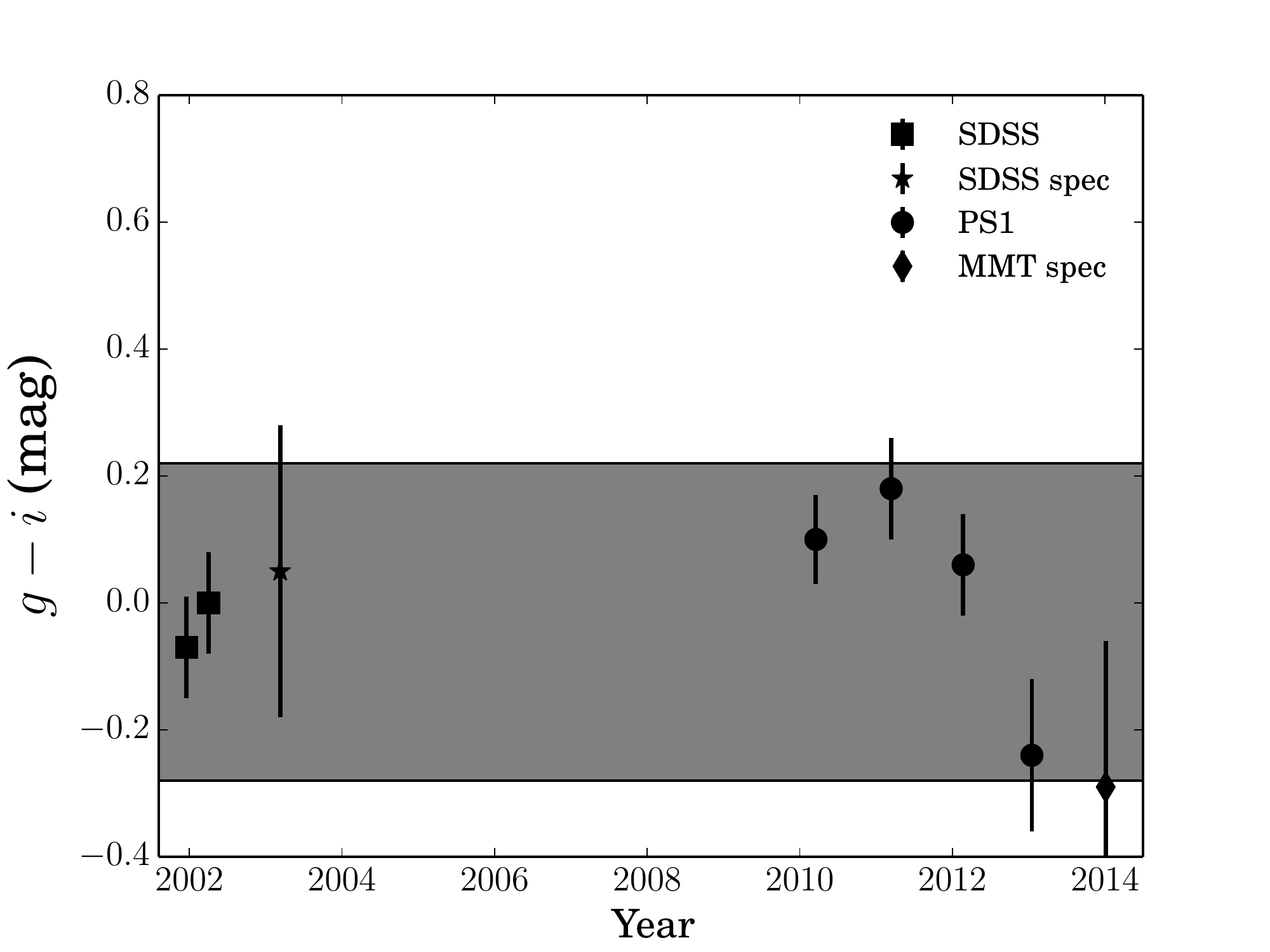}
\caption{\textit{Left}:  Measured $g$ magnitudes for SDSS1133 over the last 63 yr from DSS photographic plates, the SDSS, and PS1.  The vertical dashed line indicates a time gap.   The blue DSS filter magnitudes have been converted to $g$ magnitudes based on the SDSS colors.   The 2005 observation represents an upper limit.  We also included lower-quality synthetic photometry taken from the SDSS and MMT spectroscopy in time periods where higher-quality photometry was not available.    \textit{Right}:  Measured $g-i$ colors of SDSS1133 since its peak brightness in 2001.  Photometry is from the SDSS and PS1 during 2010--2014.  We also included lower-quality synthetic photometry taken from the SDSS and MMT spectroscopy in time periods where higher quality photometry was not available.   The grey shaded region indicates average colors of nearby ($z<0.1$) SDSS quasars \citep{Richards:2001:2308}.   }
\label{fig:bluephot}
\end{figure*}

	Imaging from multiple time periods can differentiate an AGN from a SN; the latter typically fade quickly, and they rapidly become redder in color.  Using images from the DSS, SDSS, and the PS1 survey (Fig.~\ref{fig:HistoricalImage}) along with synthetic photometry from the observed spectra, we measure the $g$ mag of SDSS1133 between 1950 and 2013 (Fig.~\ref{fig:bluephot}).  We find that SDSS1133 brightened in 2001 and 2002 compared to 1950 and 1993, and it shows variable brightening and dimming between 2010 and 2013.    A nondetection in a 2MASS image on 2000 January 10, in the NIR, suggests that significant brightening happened sometime between 2000 January 10 and the first SDSS image on 2001 December 18. SDSS1133 was detected in the near-UV at $-10.7$ mag AB in 2004 and at $-10.9$ mag AB in 2013 by the {\it GALEX} and {\it Swift} satellites (respectively), 2~yr and 12~yr after peak brightness (assumed to be the first SDSS observation on 2001 December 18).  Over the last 12~yr, the optical and UV colors of SDSS1133 have remained relatively constant (Fig.~\ref{fig:bluephot}, right) and consistent with those of quasars \citep{Richards:2001:2308}.

  \begin{figure*}
\centering
\includegraphics[width=17.2cm]{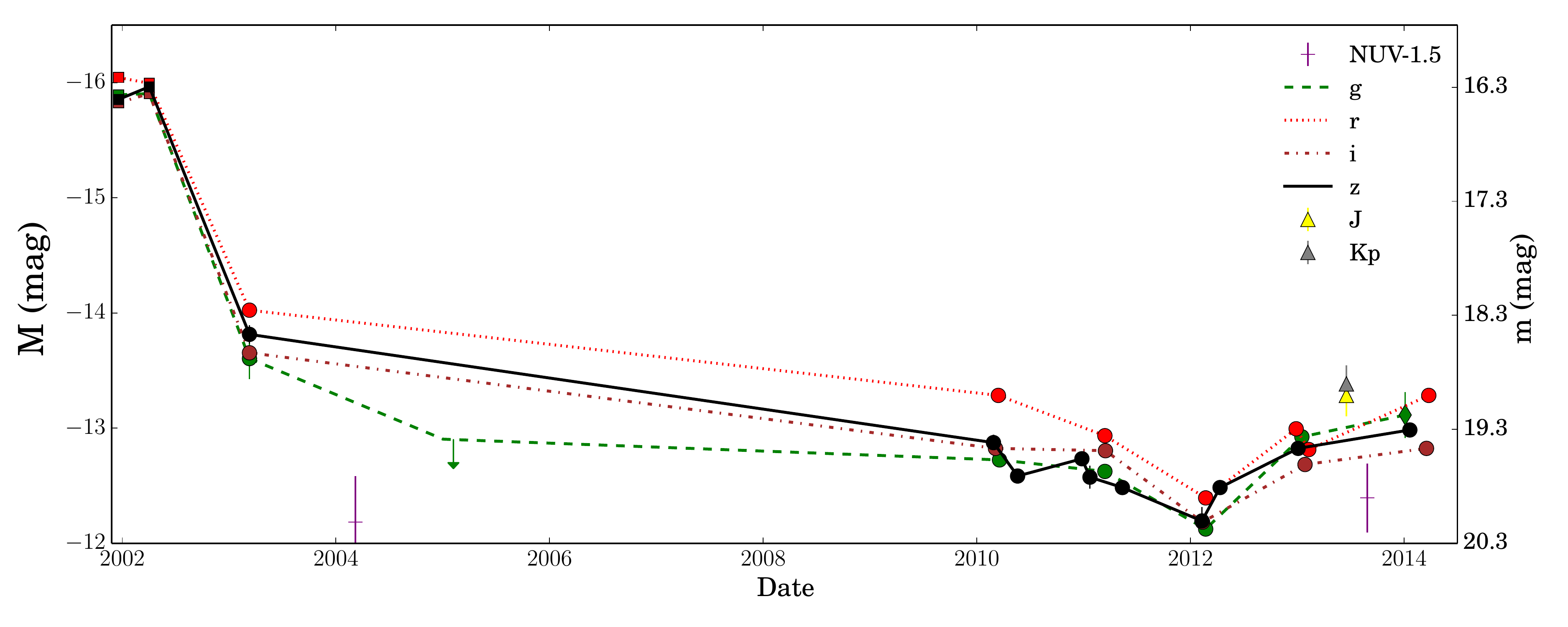}

\caption{Photometry of SDSS1133 since its peak brightness in 2001 in AB mag.   Optical observations ($griz$) are shown for the first year in 2001--2002 from the SDSS (filled squares), an upper limit from the Beijing observatory in 2005 (arrow), and finally observations from PS1 in 2010--2014 (filled circles).  Error bars are typically smaller than symbols.  We have also included lower-quality synthetic photometry taken from the SDSS and MMT (stars) in time periods where higher quality photometry was not available.  NIR AO observations from NIRC2 are shown as triangles. NUV {\it GALEX} and {\it Swift} UVOT data are shown as purple crosses and have been offset by 1.5 mag.}
 \label{fig:phot}
\end{figure*}  

\begin{figure*}
\centering
\includegraphics[width=81mm]{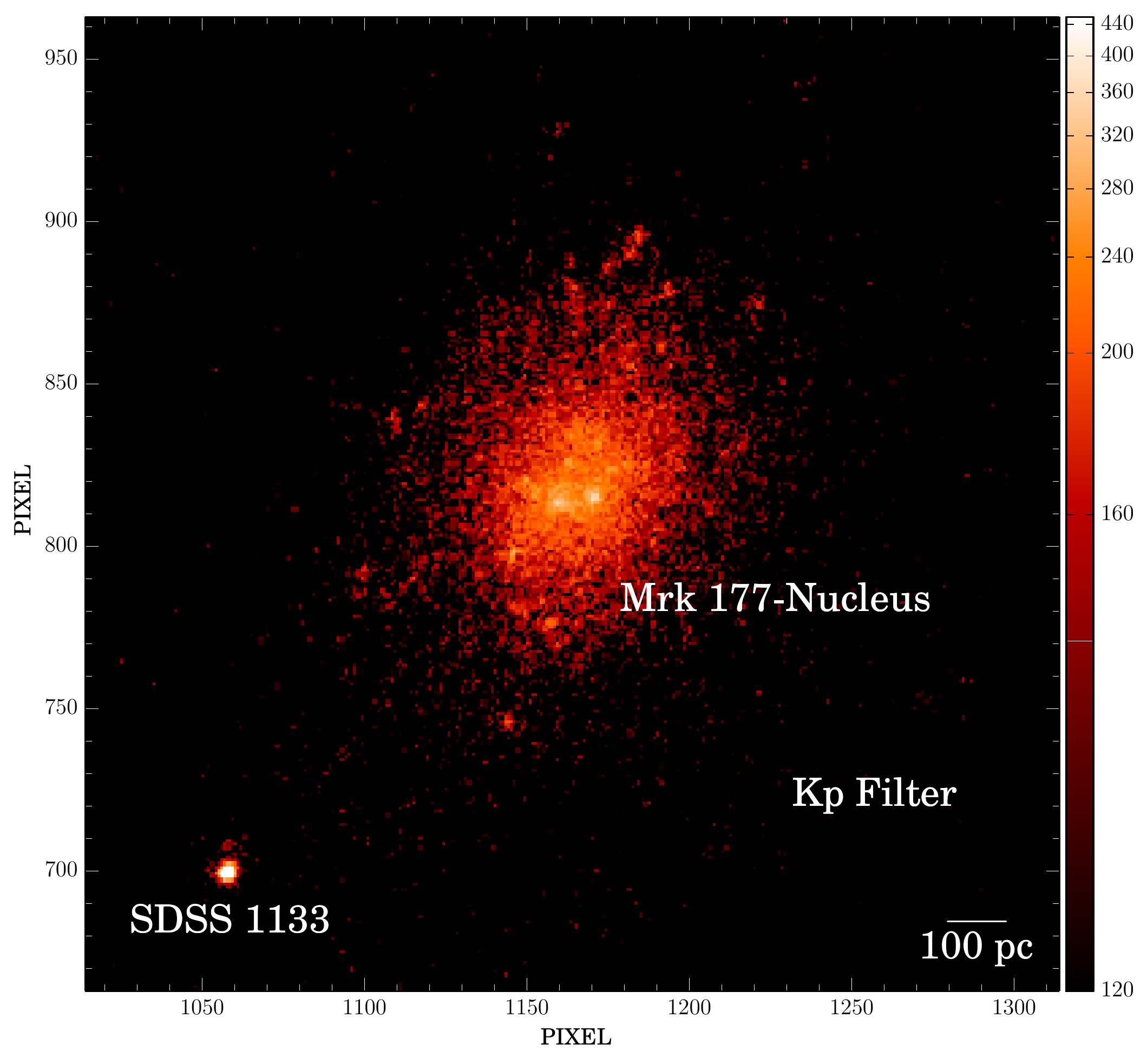}
\includegraphics[width=8.1cm]{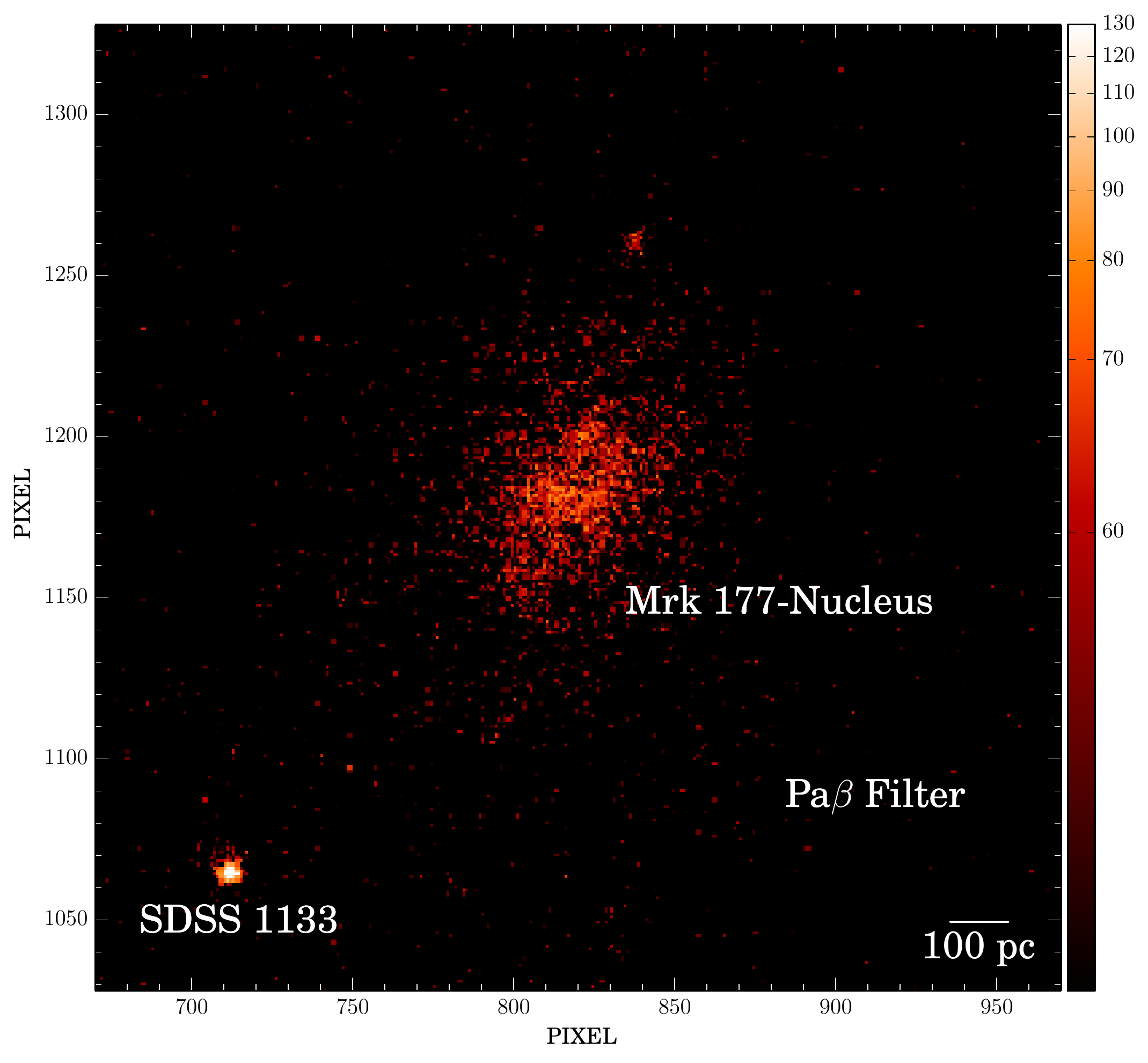}
\includegraphics[width=81mm]{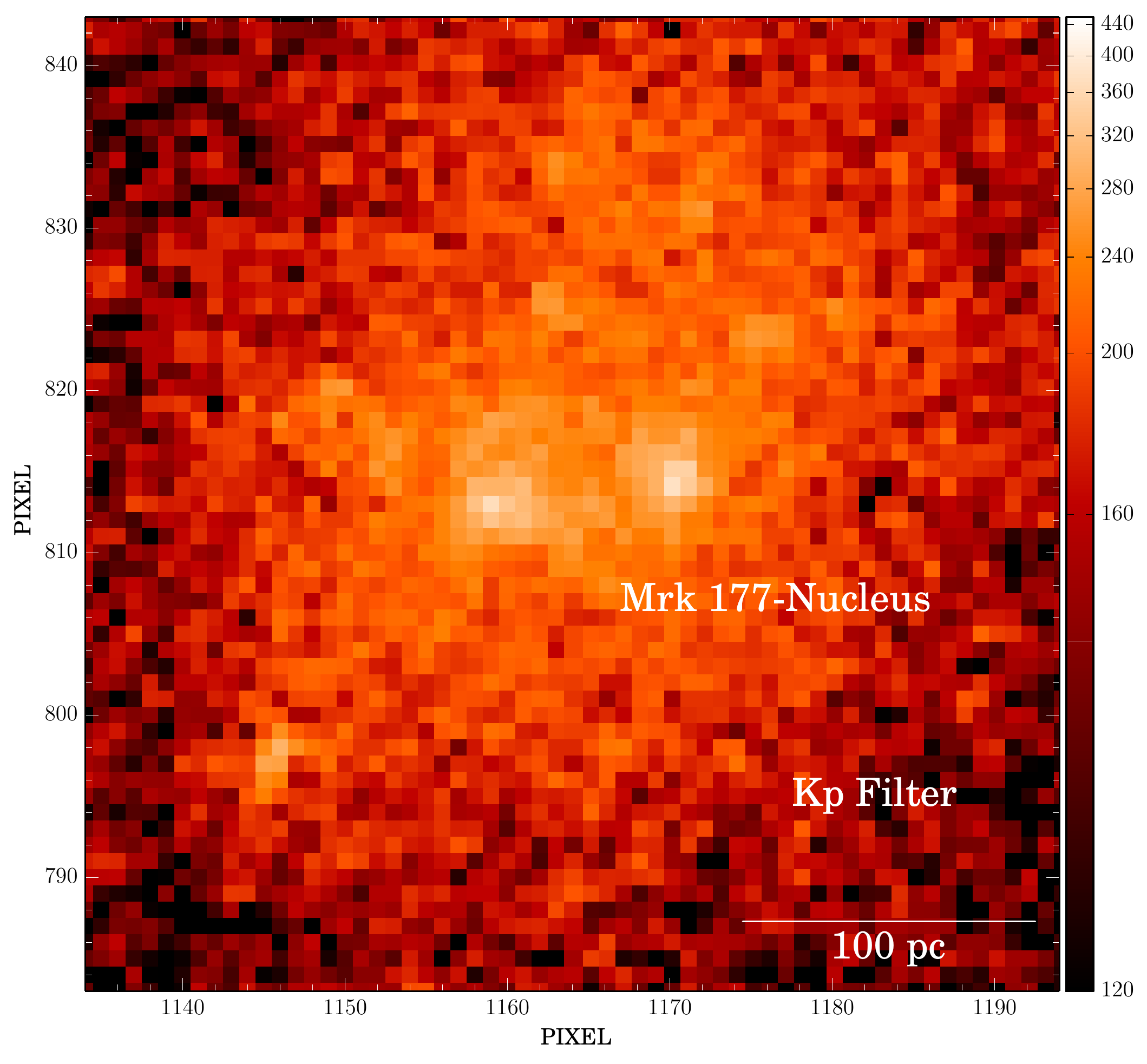}
\caption{AO images of SDSS1133 and Mrk 177.  The image is centered on Mrk 177 with a spatial scale of 40 mas pixel$^{-1}$ and displayed with an arcsinh scale in CCD counts.  \textit{Top left}: 12$\arcsec$ wide $K_p$  image.  The FWHM of SDSS1133 is consistent with the PSF image (FWHM = $0.08\arcsec$), corresponding to point-like emission at a scale of 12 pc.  \textit{Top right}:  12$\arcsec$ wide Pa$\beta$ image.   The FWHM of SDSS1133 is consistent with the PSF image (FWHM = $0.15\arcsec$) corresponding to point-like emission at a scale of 22 pc. \textit{Bottom}: 2.4$\arcsec$ $K_p$ zoomed-in image of the nuclear region of Mrk 177 showing signs of morphological disruption and two nuclei.   }
\label{fig:AO_K}
\end{figure*}

	High-resolution imaging at the $\lesssim 20$~pc scale is critical for differentiating between an ongoing merger of two galaxies and a post-merger recoiling SMBH that has left its host-galaxy nucleus behind.  We imaged a 40$\arcsec$ region around SDSS1133 in the NIR Pa$\beta$ and $K_p$ filters using AO with the Near Infrared Camera 2 (NIRC2) instrument on the Keck-II telescope (Fig.~\ref{fig:AO_K}).  The images reveal an unresolved point source at spatial scales of $\lesssim 12$ ($K_p$) and $\lesssim 22$~pc (Pa$\beta$) coincident with the location of SDSS1133.  SDSS1133 shows average colors ($g-i \approx -0.3$ to 0.2 mag) constant within the uncertainties from 2002 through 2014.  This color is consistent with nearby ($z<0.1$) SDSS quasars \citep{Richards:2001:2308}.  Additionally, the $i_{\rm AB}-K_{\rm Vega}$ color of SDSS1133 is 2.5 mag, in agreement with results for low-redshift quasars \citep{Peth:2011:105} and much redder than the stellar locus ($i-K \approx 1$ mag).  While red colors can be found in SNe because of dust formation, these colors often evolve with time.
	
\begin{figure*}
\centering
\includegraphics[width=8.1cm]{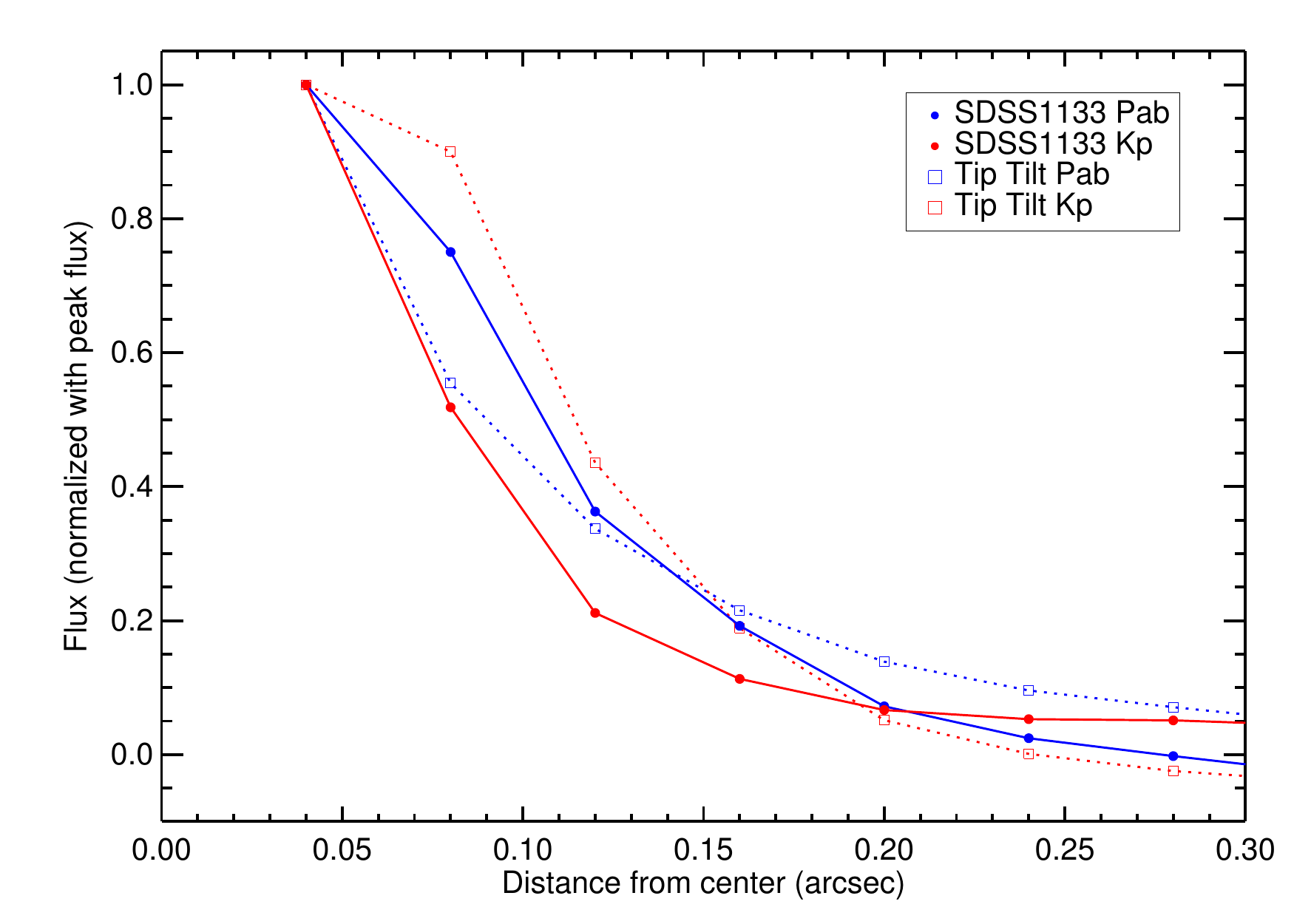}
\caption{AO radial profiles of SDSS1133 (solid line) and the tip-tilt star (dashed line) used for PSF estimation in each observation.  0.1$\arcsec$ corresponds to a physical scale of 14 pc.  The radial profiles of the $K_p$ (FWHM = $0.08\arcsec$) and Pa$\beta$ (FWHM = $0.15\arcsec$) of SDSS1133 images are consistent with the stellar profiles of the tip-tilt stars.       }
\label{fig:AO_radprof}
\end{figure*}

	NIR images trace the old stellar population and the bulk of the stellar mass. Those of SDSS1133 exhibit a disturbed morphology and a double-nucleus structure with a separation of 60~pc. The $K_p$ absolute magnitudes of the nuclei are $-14.1$ and $-14.2$, in the range of dwarf-galaxy nuclei or very young, massive star clusters \citep{Forbes:2008:1924}.  Radial profiles of SDSS1133 and the tip-tilt star used in each observation are shown in Figure~\ref{fig:AO_radprof}.  The radial profiles of the nuclei in $K_p$ and Pa$\beta$ are consistent with the stellar profiles of the tip-tilt stars, suggesting no extended emission.

        With the {\it Swift} XRT at 0.3--10 keV, $7.6 \pm 3.4$ background-subtracted counts were detected, corresponding to a signal-to-noise ratio (S/N) of 2.2 and a net count rate of $(4.1 \pm 1.8) \times 10^{-4}$ counts s$^{-1}$.  The marginal {\it Swift} X-ray detection cannot distinguish between the AGN and SN scenarios.  Assuming a power-law index of 1.9, representative of an AGN, the luminosity is $1.5\times10^{39}$ $\ergps$ in the 0.3--10 keV band.

\subsection{Spectral Shape, Narrow-Line Diagnostics, Broad-Line Emission}
	We measure the amount of narrow-line emission detected in SDSS1133, the nucleus of the host galaxy Mrk 177, and at a position the same radial distance from the nucleus of Mrk 177 as SDSS1133 (Mrk 177 IFU-offset) using the IFU image; the latter allows us to determine the expected level of narrow-line contamination in SDSS1133 from Mrk 177.   All line offsets are from the wavelength of [O~III] $\lambda$5007 in Mrk 177 at $z=0.007845$.  The measured line offset of the narrow lines in SDSS1133 and the host galaxy, Mrk 177, is redshifted by a small amount ($26 \pm 9~\kmps$), while Mrk 177 IFU-offset is blueshifted by a small amount ($-18 \pm 12~\kmps$).  The narrow-line FWHMs of SDSS1133 and Mrk 177 are consistent with the SDSS instrumental resolution (150 $\kmps$).  The FWHM of Mrk 177 IFU-offset is consistent with the spectral resolution of the IFU (360 $\kmps$). 

\begin{figure*}
\centering
\includegraphics[width=5.4cm]{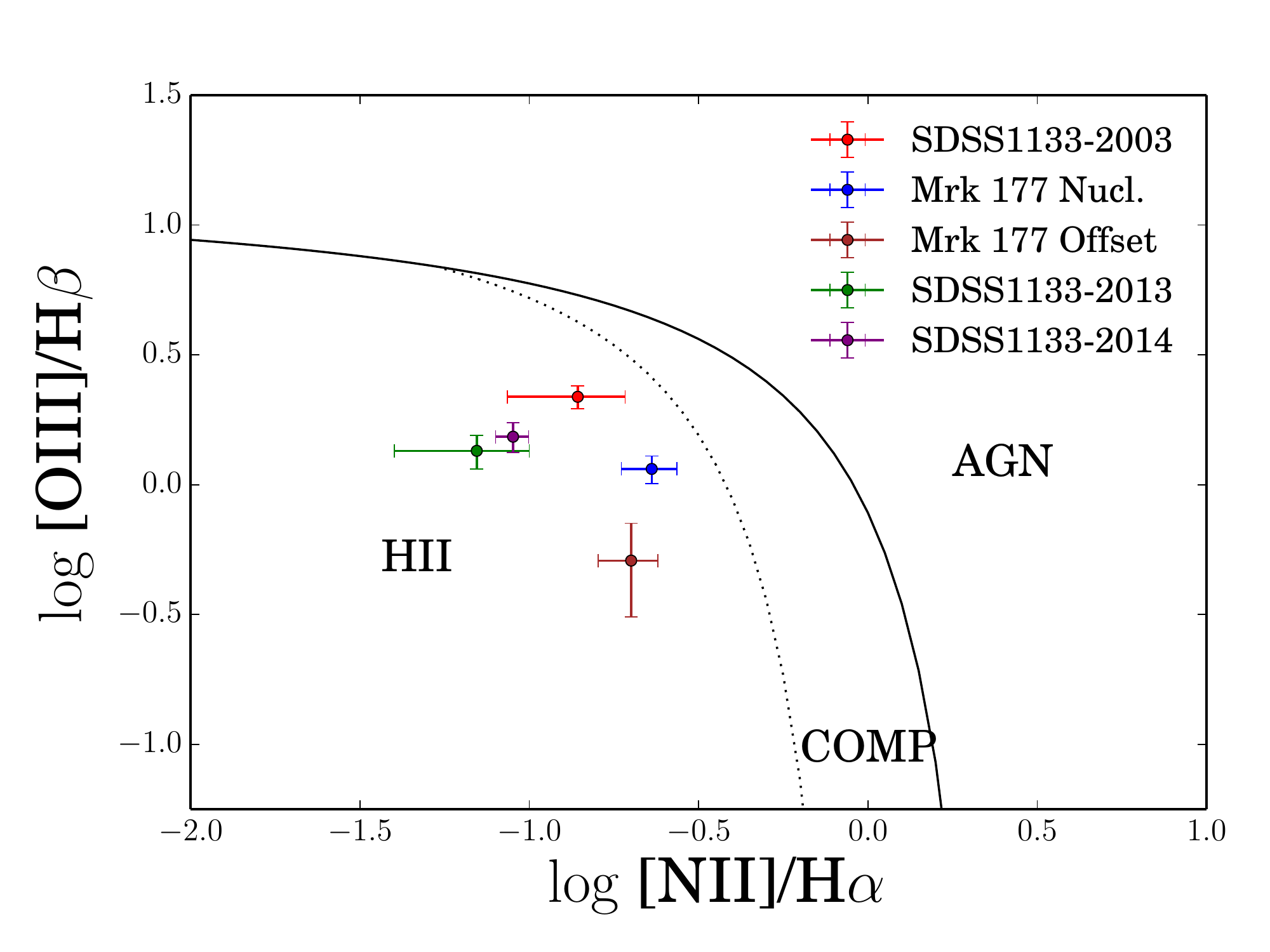} 
\includegraphics[width=5.4cm]{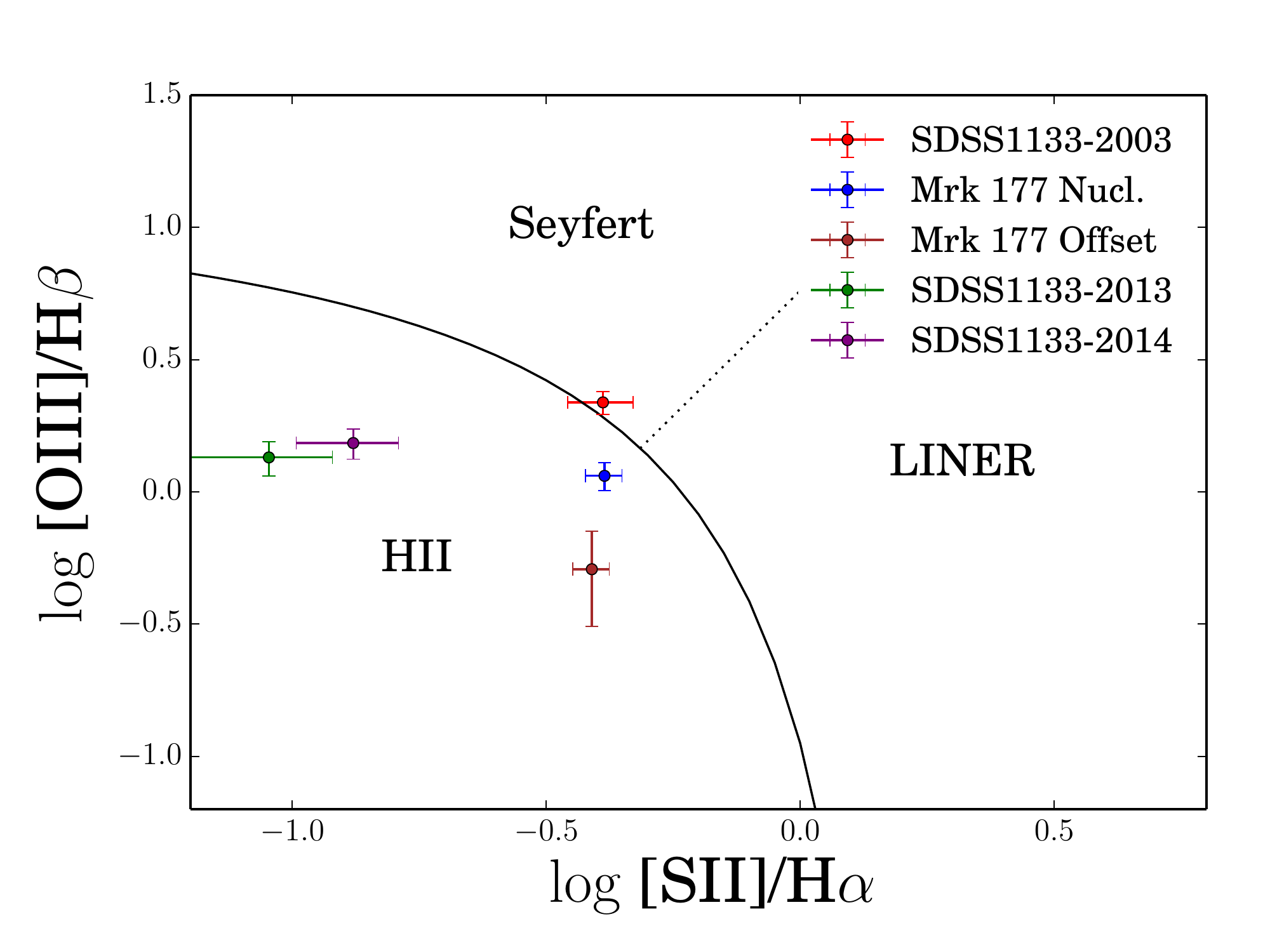} 
\includegraphics[width=5.4cm]{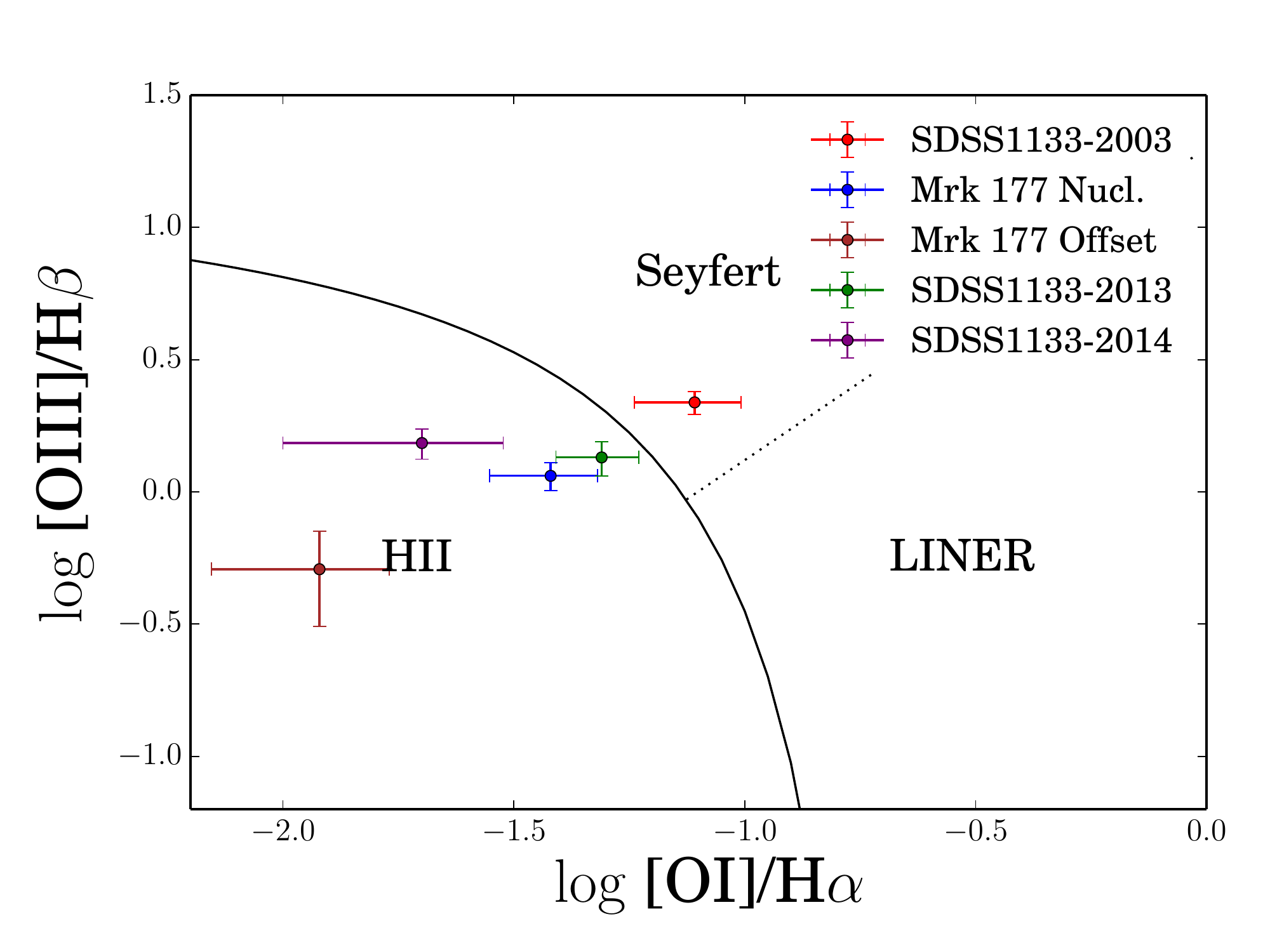}
\caption{Optical diagnostic diagrams [N~II]/H$\alpha$,  [S~II]/H$\alpha$, and [O~I]/H$\alpha$ vs. [O~III]/H$\beta$ \citep[e.g.,][]{Kewley:2006:961} for SDSS1133 and the host galaxy Mrk 177 from the SDSS spectra.  Lines indicate the approximate separation between H~II regions, composite regions, Seyfert AGNs, and LINER AGNs.   }
\label{fig:Narrow_Emission}
\end{figure*}

	We apply AGN emission-line diagnostics \citep[e.g.,][]{Kewley:2006:961,Veilleux:1987:295} to the [N~II]/H$\alpha$, [S~II]/H$\alpha$, [O~I]/H$\alpha$ narrow lines of SDSS1133 and Mrk 177 (Fig.~\ref{fig:Narrow_Emission}).  The host galaxy, Mrk 177, is classified as an H~II region in all diagnostics. SDSS1133 is classified as a Seyfert with the [S~II] and [O~I] diagnostics in 2003 SDSS spectra.

	We find that the [N~II] luminosity of the narrow-line emission found in SDSS1133 in 2003 and 2013 is consistent with that of Mrk 177 IFU-offset.  However, the [O~III], H$\beta$, H$\alpha$, and [O~II] lines are significantly stronger than in Mrk 177 IFU-offset.  This is confirmed by the two-dimensional images, which show that the emission lines are above the continuum level of SDSS1133 and Mrk 177.  By 2013, the [O~III] luminosity decreased.

\begin{figure*}
\includegraphics[width=5.4cm]{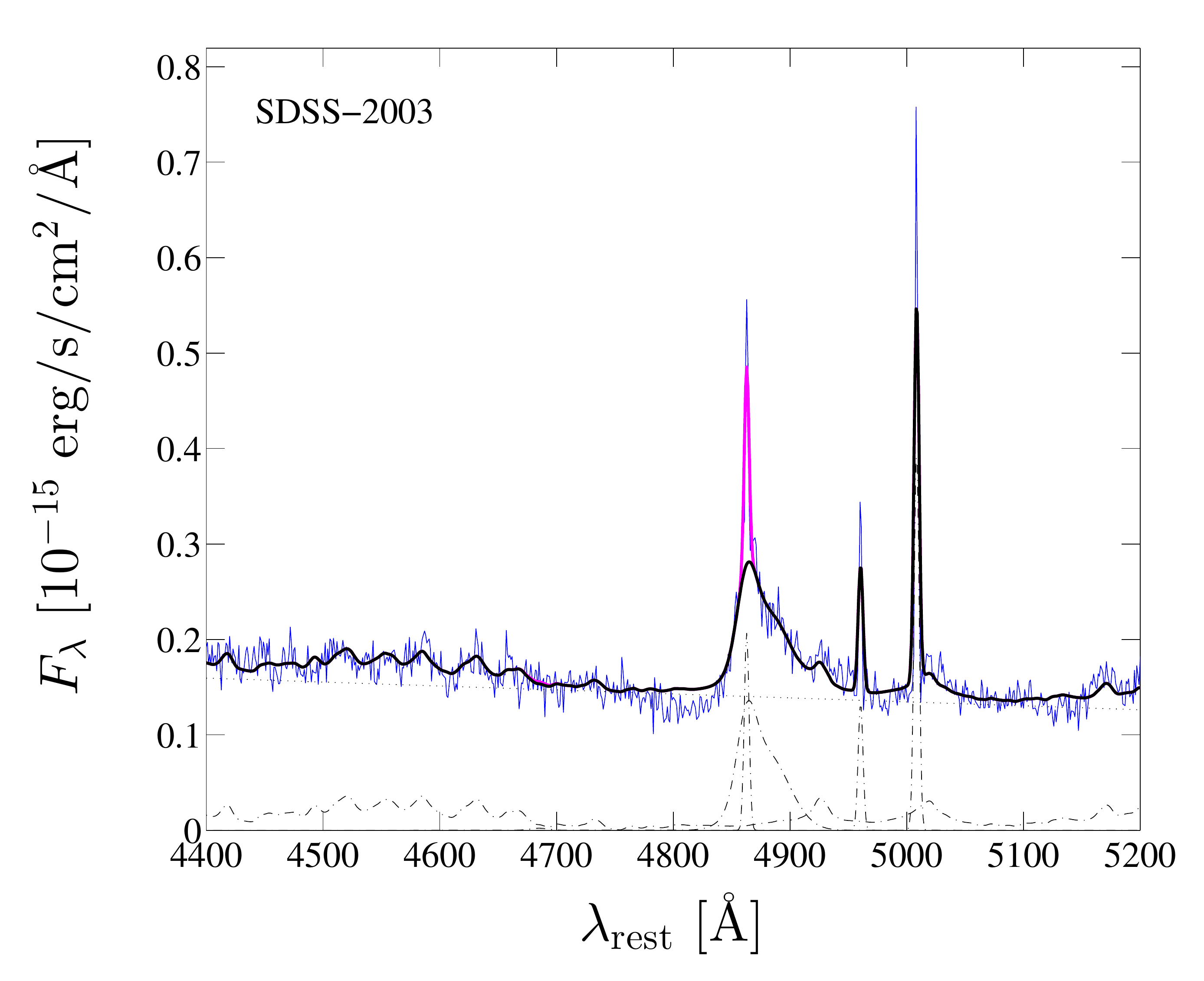} 
\includegraphics[width=5.4cm]{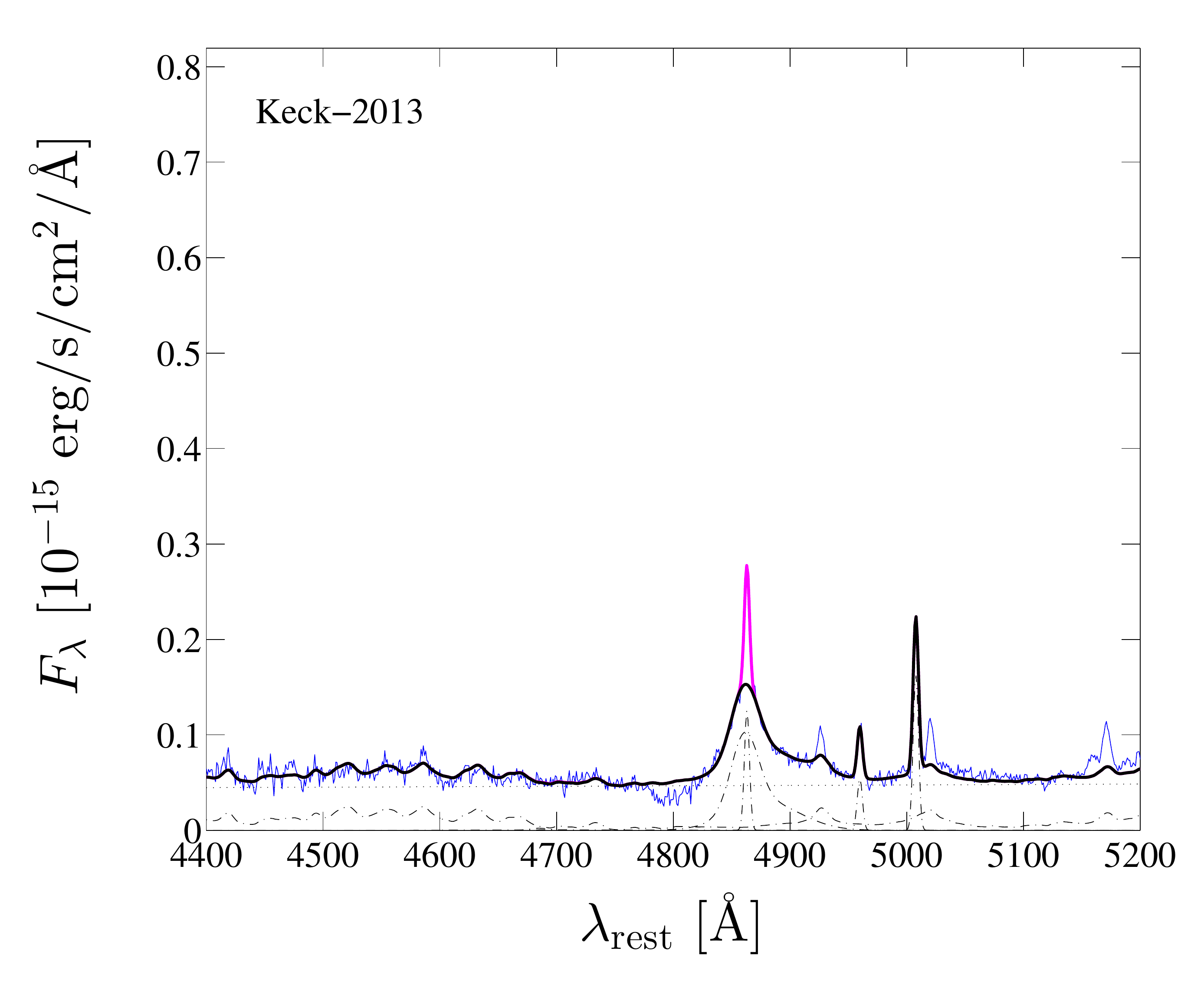} 
\includegraphics[width=5.4cm]{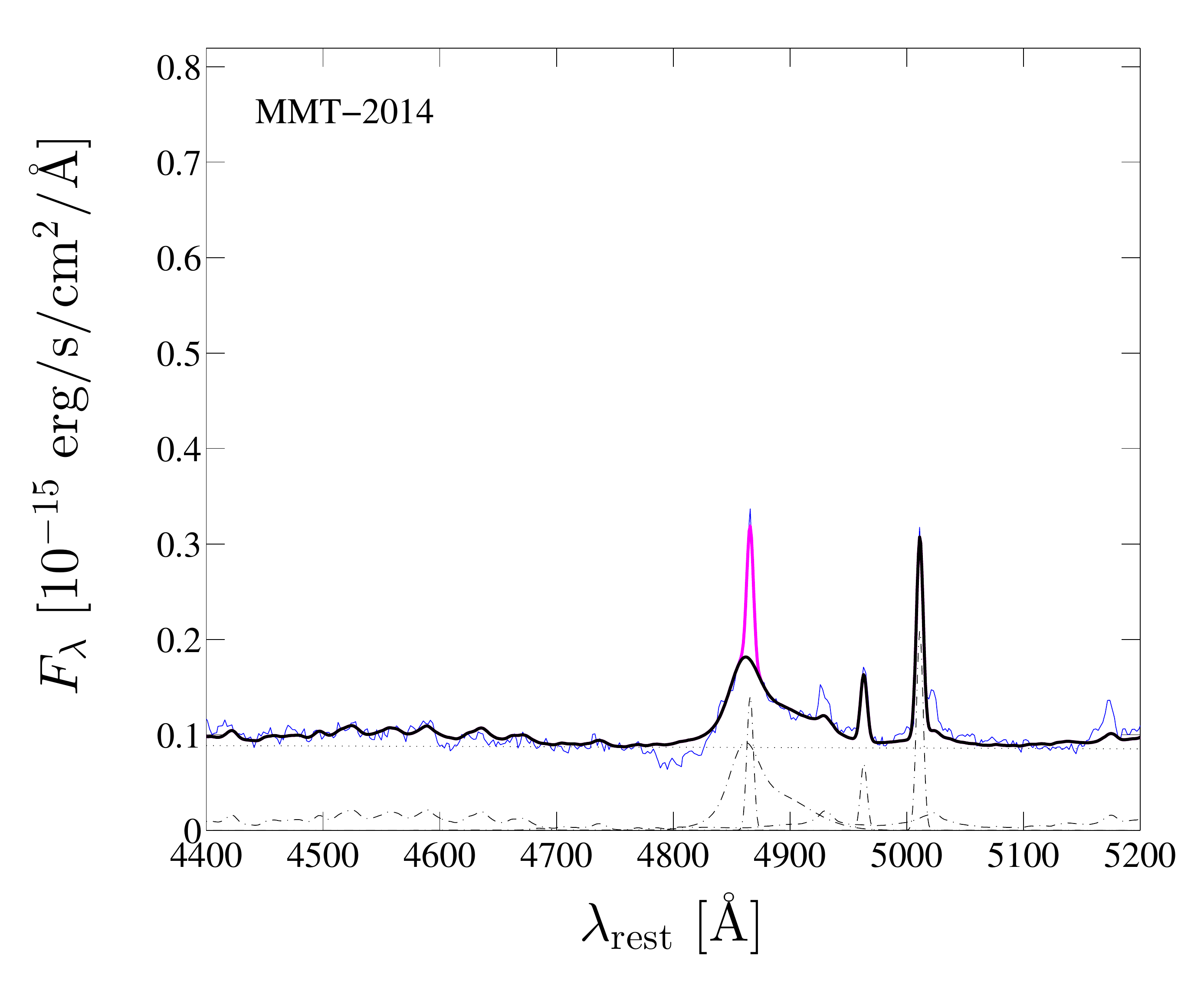}
\caption{\textit{Left}: Fits to the H$\beta$ region of SDSS1133 from 2003 (SDSS, left), 2013 (Keck, middle), and 2014 (MMT, right).  The 2013 and 2014 spectra show narrow Fe~II emission, weaker continuum emission, and a lower ratio of [O~III] to H$\beta$ compared to the 2003 spectrum.        }
\label{fig:Hbeta}
\end{figure*}

\begin{figure*} 
\centering
\includegraphics[width=5.4cm]{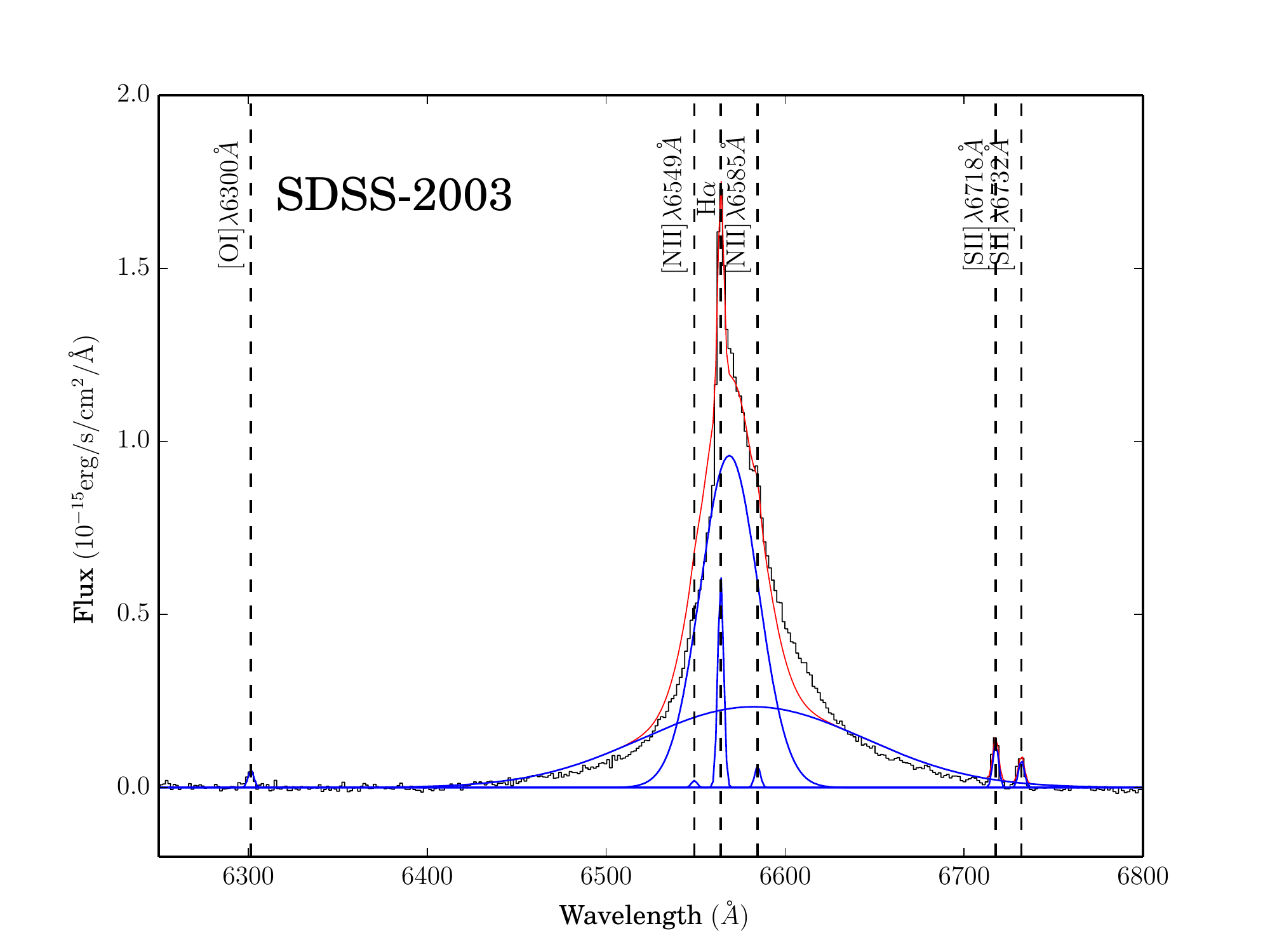}
\includegraphics[width=5.4cm]{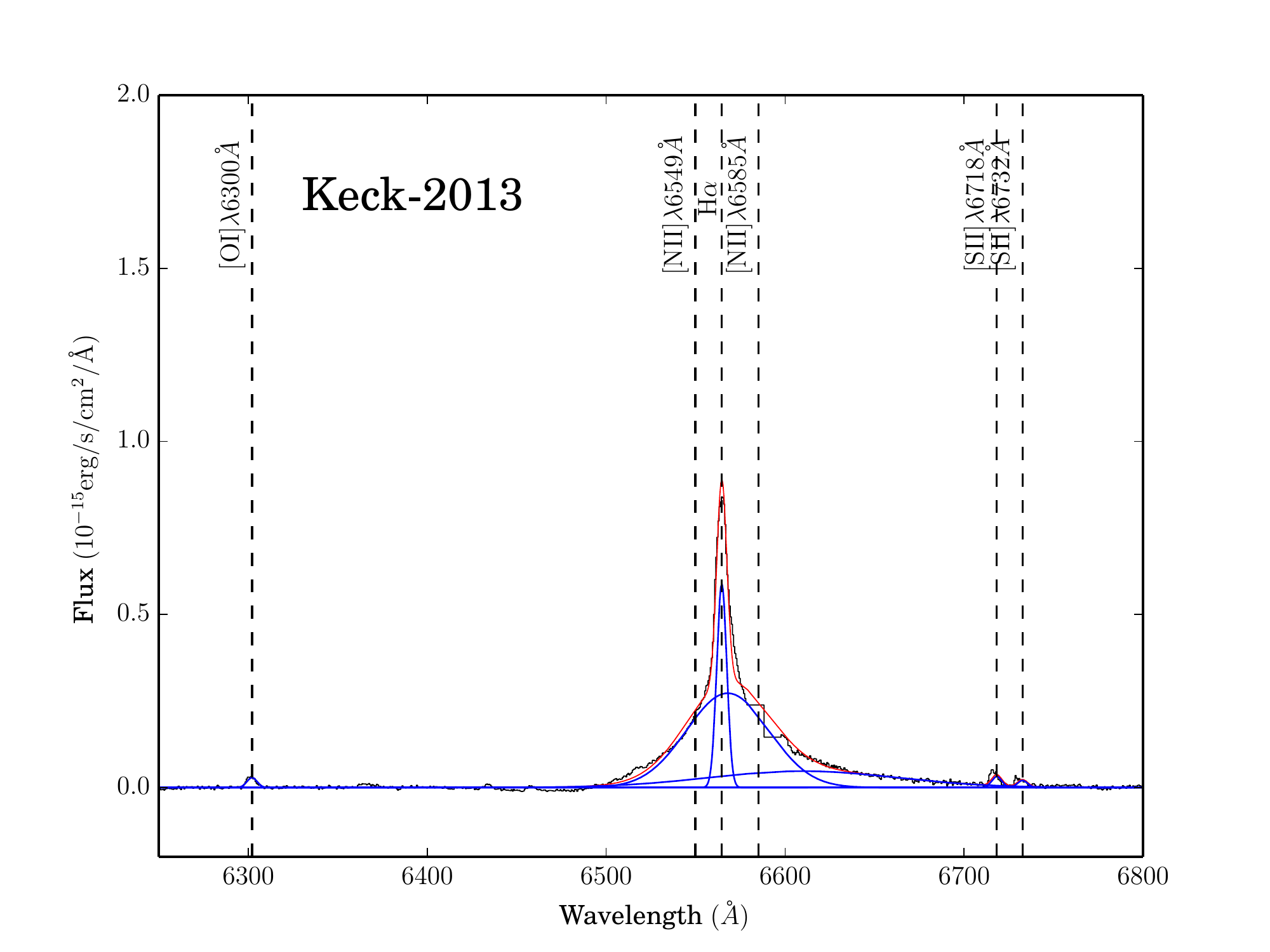}
\includegraphics[width=5.4cm]{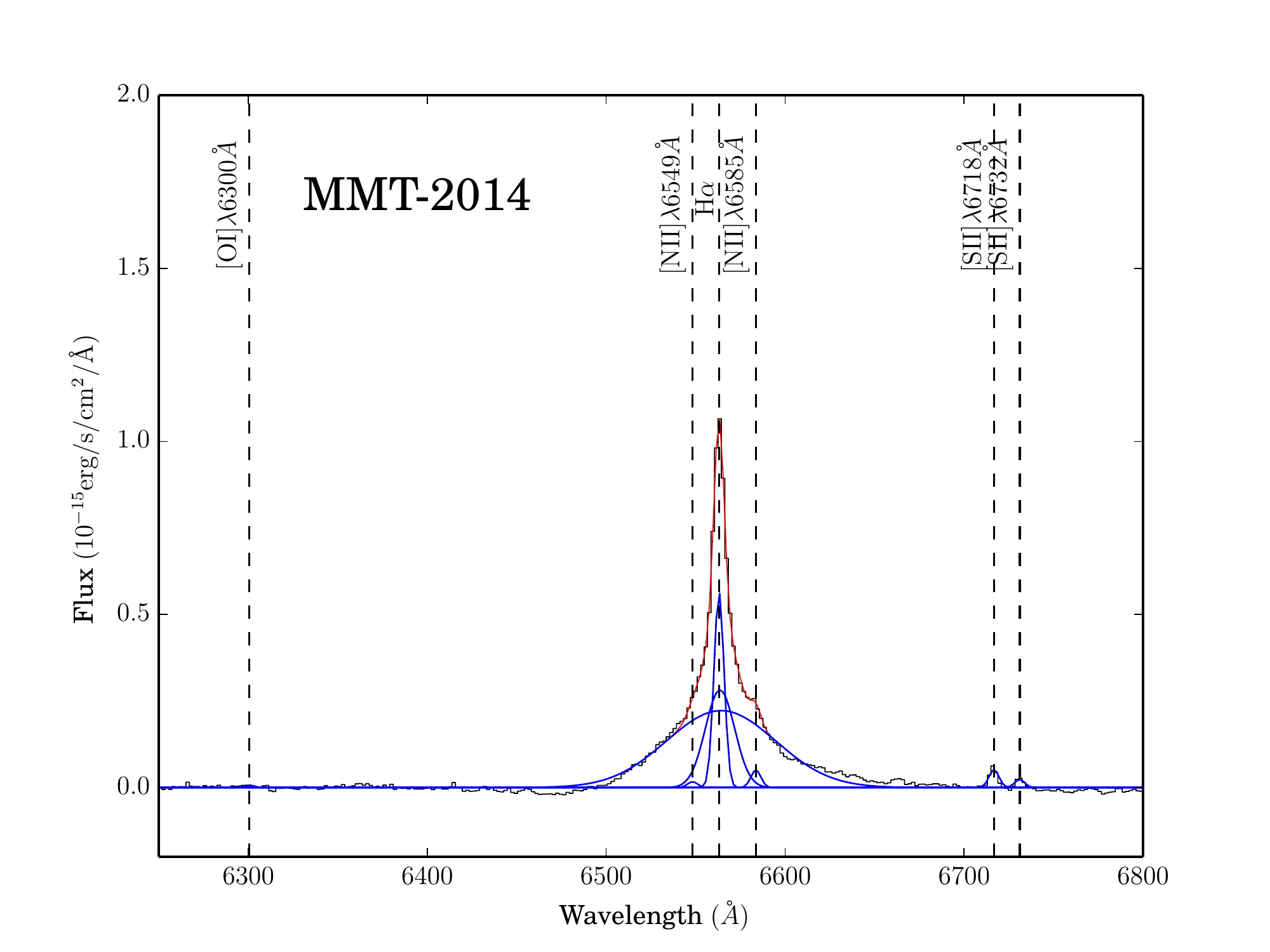}
\caption{H$\alpha$ region of SDSS1133 from 2003 (SDSS, left), 2013 (Keck, middle), and 2014 (MMT, right).  A power-law continuum has been subtracted from the spectra.  Black lines indicate the observed spectra, blue lines are model components, and red lines represent the summed model spectra.  Emission lines are shown at the redshift of the host galaxy Mrk 177.   }
\label{fig:Halpha}
\end{figure*}

	A plot of the H$\beta$ spectral region can be found in Figure~\ref{fig:Hbeta}.  We see that the 2013 Keck spectrum exhibits narrow Fe~II emission, less-broad H$\beta$ emission (2020 $\kmps$ vs.~2530 $\kmps$), lower luminosity, weaker continuum emission, and a lower intensity ratio of [O~III] to H$\beta$ compared to the 2003 spectrum.  Examining H$\alpha$ in 2013 (Fig.~\ref{fig:Halpha}), we find wider broad H$\alpha$ emission (1390$\kmps$ vs.~1660 $\kmps$) and lower luminosity than in 2003.  Assuming SDSS1133 is an AGN, we fit the H$\beta$ region and measure a SMBH mass \citep{Grier:2013:90} of  $\log\left(M_{\rm BH} / M_{\odot}\right) \approx 6.0$, 6.0, and 6.2 in 2003, 2013, and 2014, respectively.  Like all single-epoch (``virial") $M_{\rm BH}$ determinations, these estimated values have uncertainties of about 0.5 dex \citep{Shen:2013:61}, suggesting that the measured SMBH masses are constant within the uncertainties.
	
\begin{figure*} 
\centering
\includegraphics[width=8.1cm]{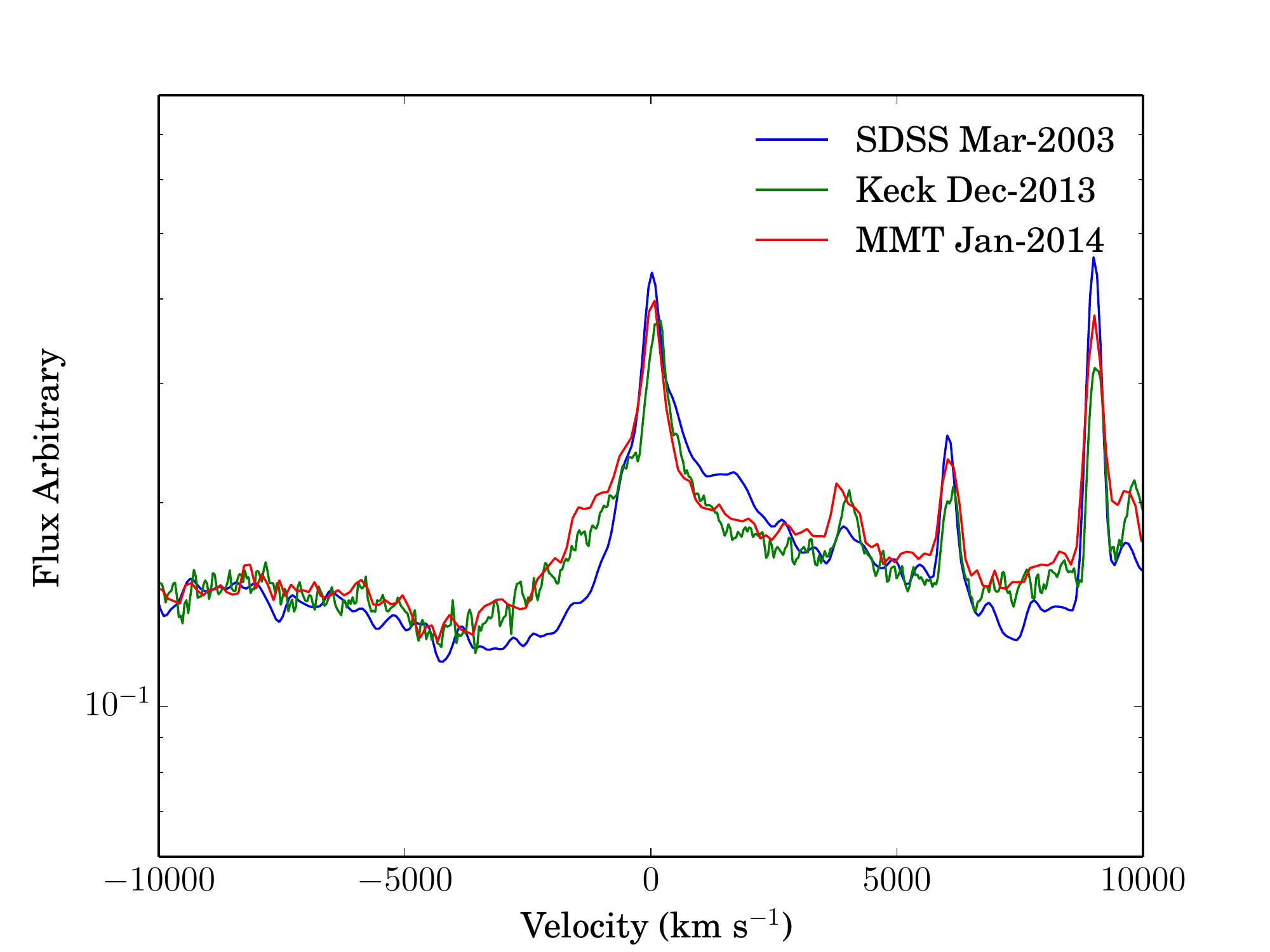}
\includegraphics[width=8.1cm]{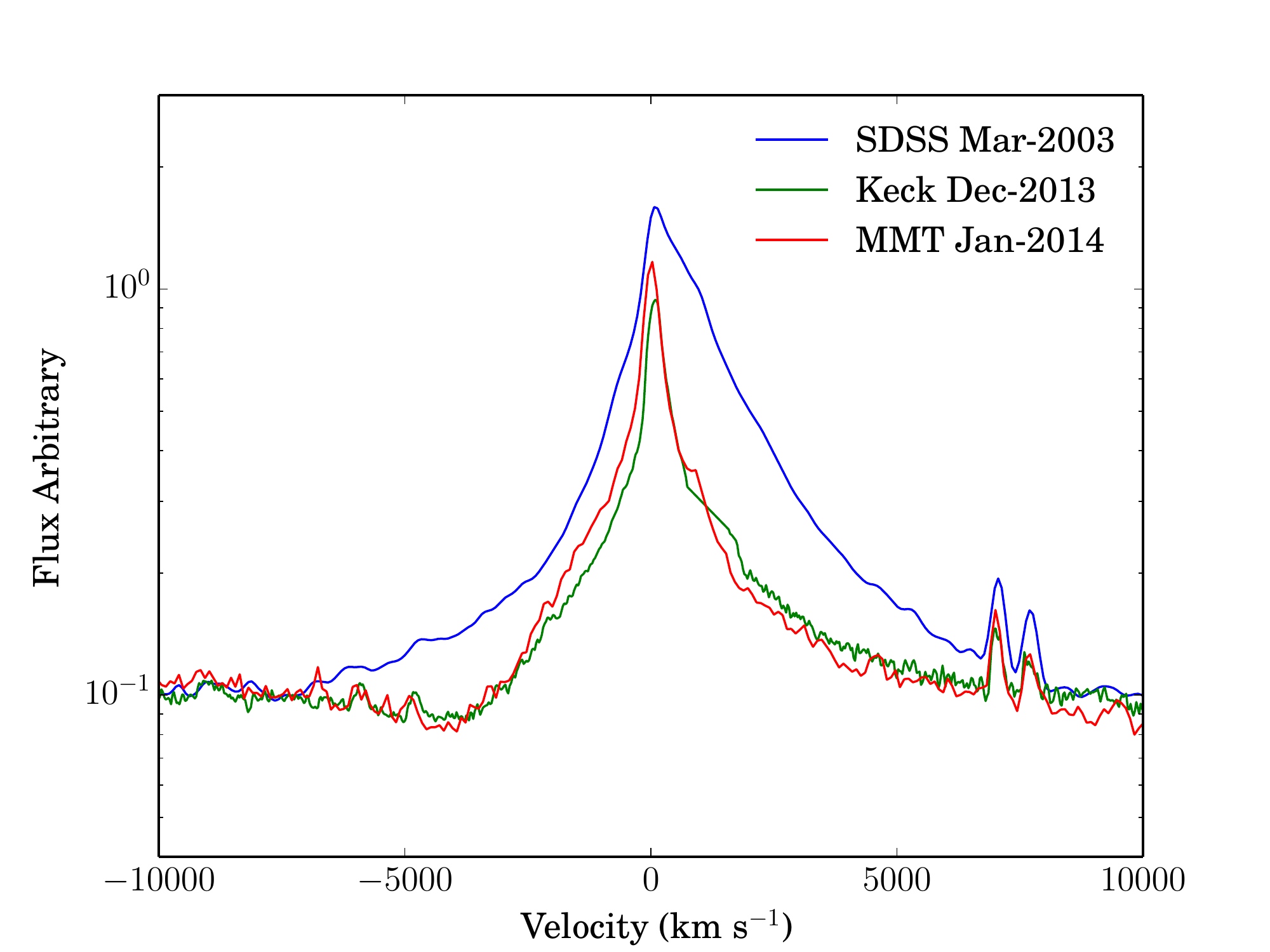}
\caption{Broad-line region of SDSS1133 in H$\beta$ (left) and H$\alpha$ (right) from SDSS in 2003, Keck in 2013, and the MMT in 2014.  The spectra have been smoothed to the same spectral resolution, offset to the same continuum level, and plotted on a log scale in the ordinate.  There is evidence of blueshifed absorption in the H$\beta$ and H$\alpha$ regions at $-3000$ to $-8000~\kmps$ in almost all of the spectra except the 2003 SDSS spectra. The total broad H$\beta$ emission dropped by 10--20\% between 2003 and 2013--2014, whereas the total broad H$\alpha$ emission dropped by 65--70\%.
}
\label{fig:pcygni}
\end{figure*}

	A plot of the broad Balmer lines smoothed to the same resolution and offset to the same continuum can be found in Figure~\ref{fig:pcygni}.  There is evidence of blueshifted absorption in the H$\beta$ and H$\alpha$ regions at $-3000$ to $-8000$ $\kmps$. The total broad H$\beta$ emission dropped by 10--20\% between 2003 and 2013--2014, whereas the total broad H$\alpha$ emission dropped by 65--70\%.  The broad H$\alpha$/H$\beta$ intensity ratio is 13 in 2003 and 5 in 2013--2014. The Balmer decrement in 2003 for the broad hydrogen lines is considerably larger than the recombination value, suggesting that the broad emission originates in very dense gas or suffers higher extinction.  Finally, the change in blueshifted absorption, which is different in 2003 than in 2013--2014, may cause variation in the broad-line emission and Balmer decrement.

\begin{figure*} 
\centering
\includegraphics[width=4.2cm]{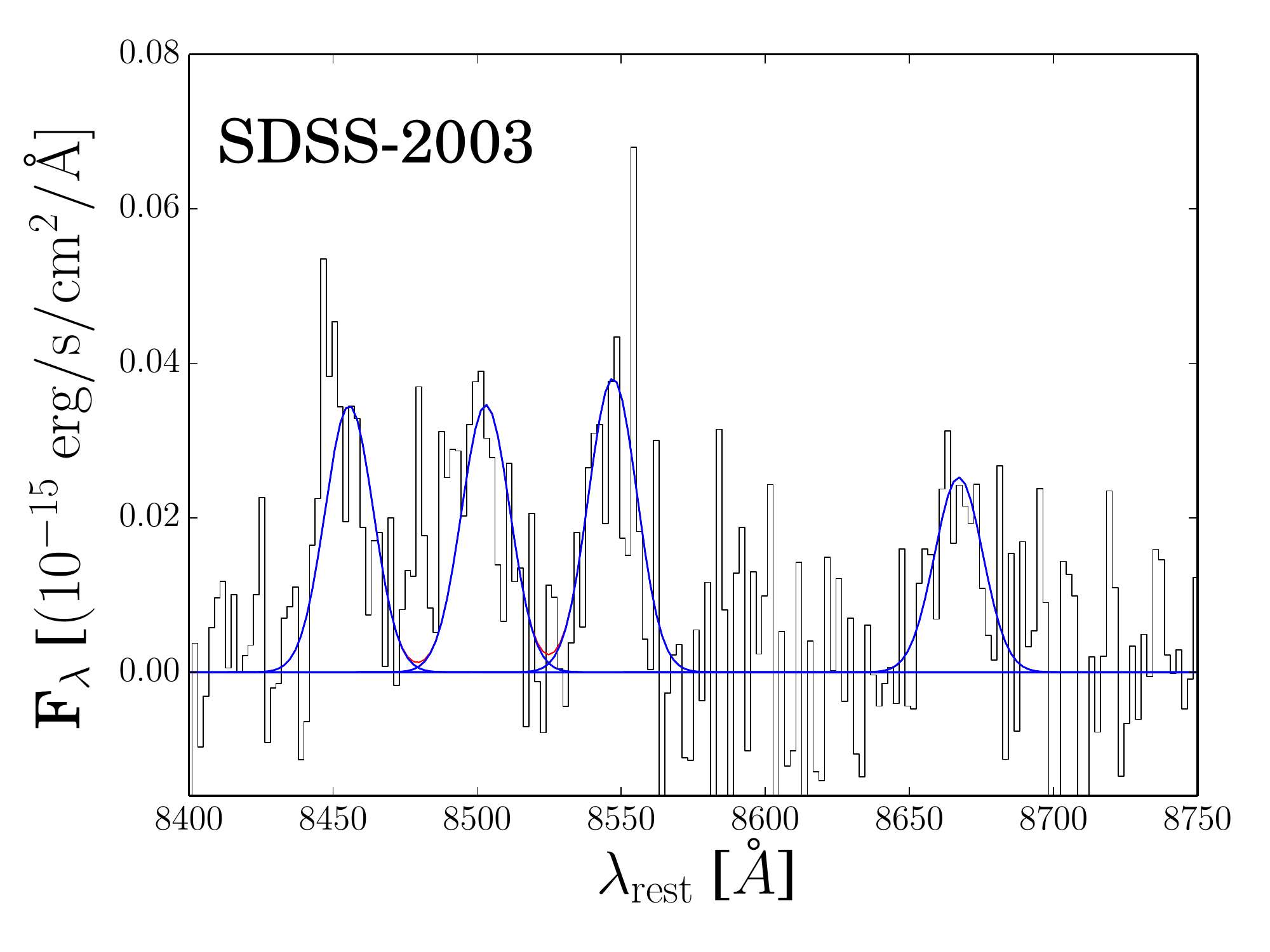}
\includegraphics[width=4.2cm]{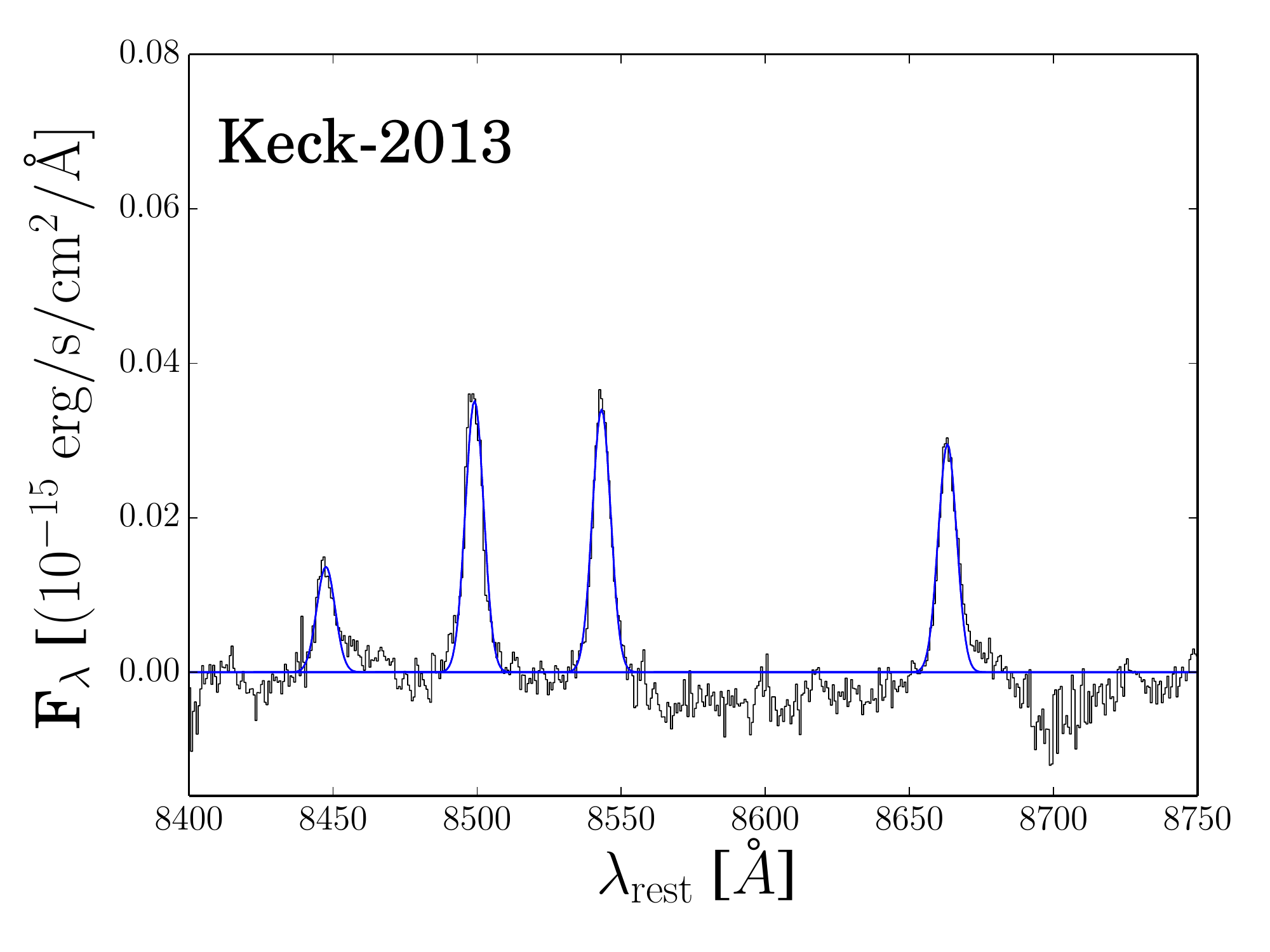}
\includegraphics[width=4.2cm]{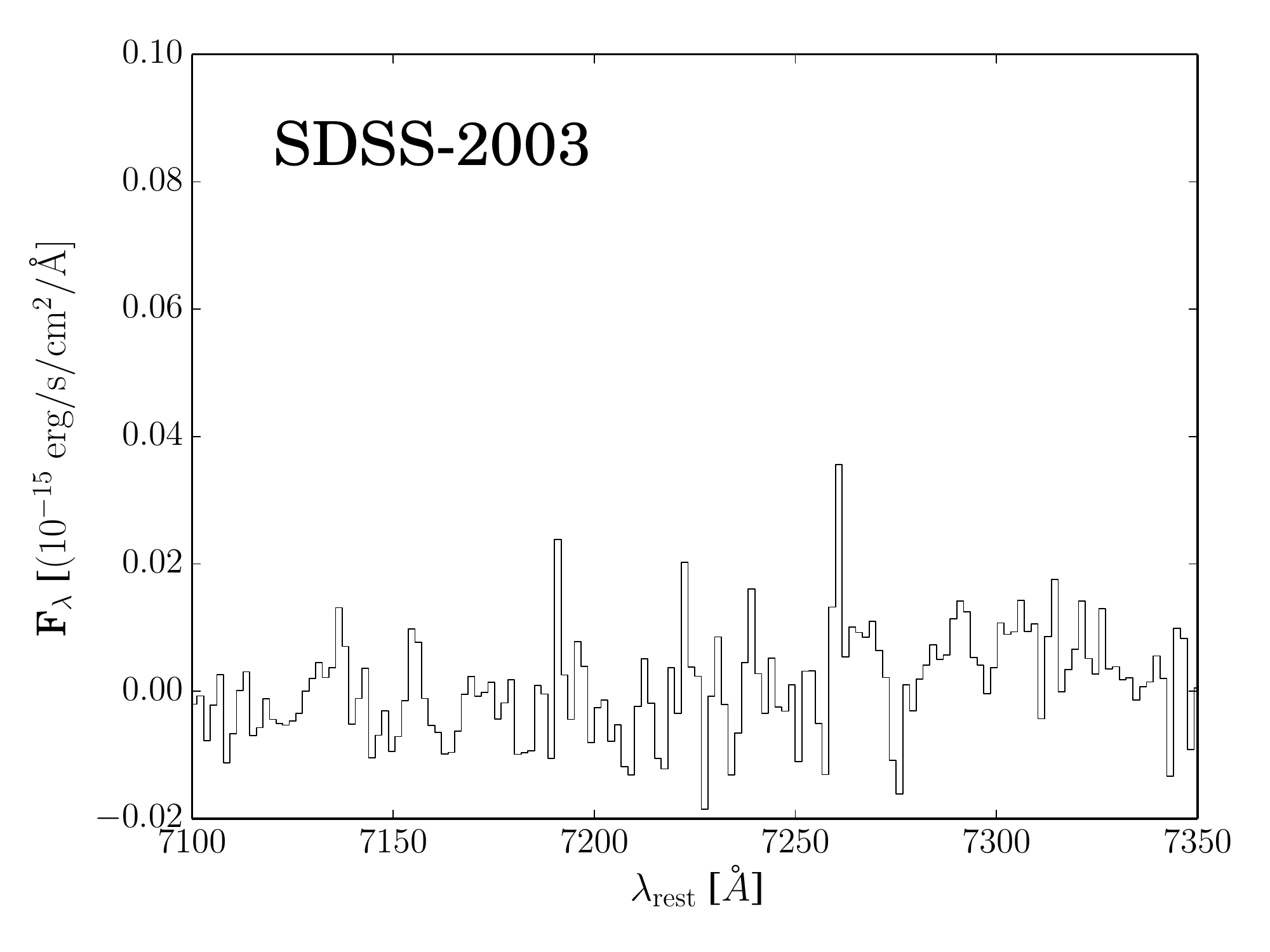}
\includegraphics[width=4.2cm]{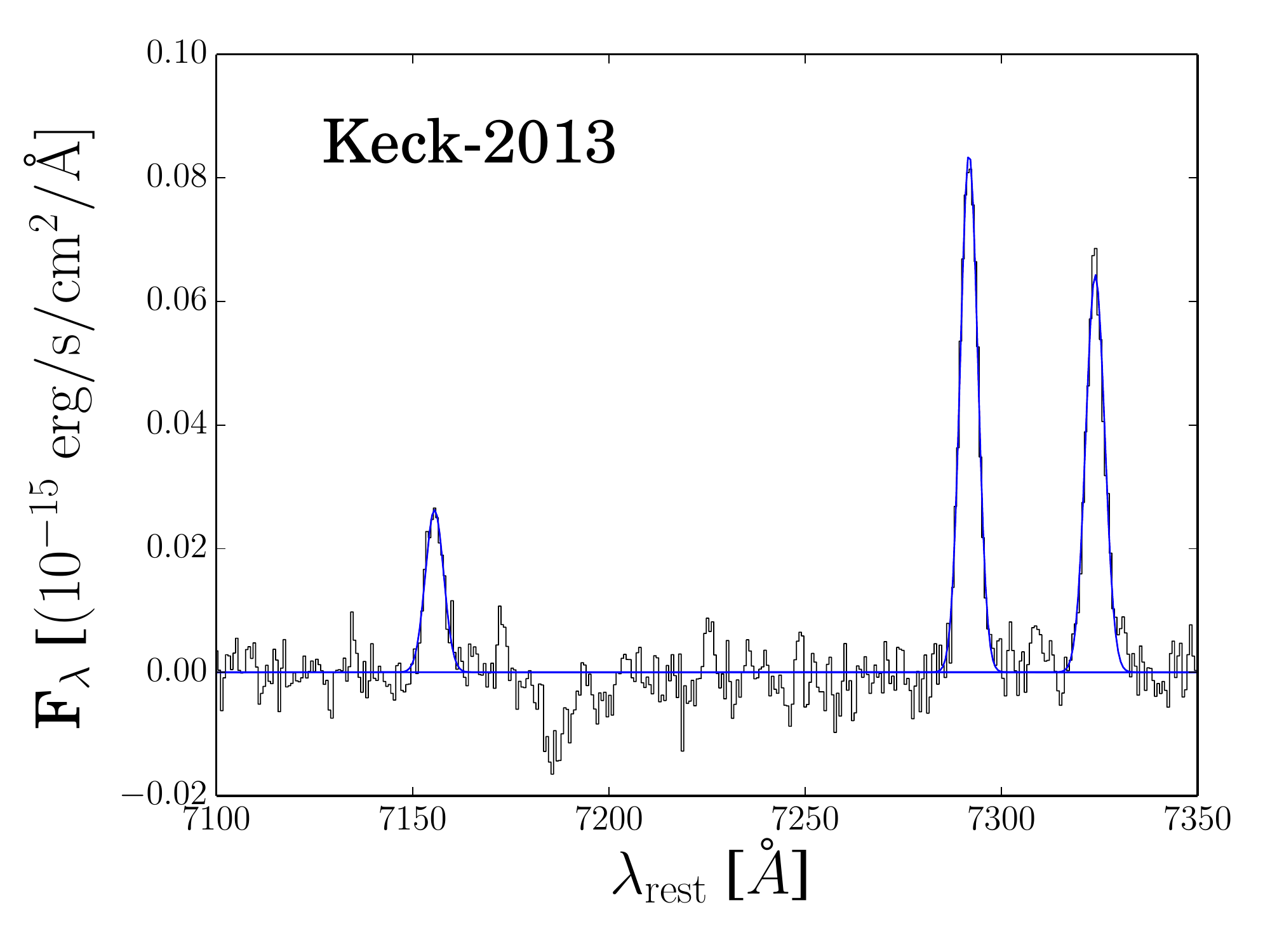}
\caption{Calcium near-IR triplet region of SDSS1133 as observed in 2003 (left) and 2013 (middle left), showing line fits to O~I $\lambda$8446 and Ca~II $\lambda\lambda$8498, 8452, 8662.  The 2003 spectra shows broader emission.  Fits to [Fe~II] $\lambda$7155 and [Ca~II] $\lambda\lambda$7291, 7324 in 2003 (middle right) and 2013 (right) are also shown.  No emission lines are detected in 2003, whereas the 2013 spectrum exhibits strong, narrow emission consistent with the resolution of the instrument (220 $\kmps$).    }
\label{fig:Catrip}
\end{figure*}

       A feature associated with the broad-line region in 1/3 of quasars \citep{Netzer:1990:57}  and some SNe is broad emission in the Ca~II NIR triplet ($\lambda\lambda$8498, 8452, 8662) and O~I $\lambda$8446 (Fig.~\ref{fig:Catrip}). In 2003, we find the FWHM of these lines to be 630--650 $\kmps$, significantly higher than that of the narrow-line region, and offset by $117 \pm 36$ $\kmps$ for the Ca~II triplet and $210 \pm 52$ $\kmps$ for O~I.   The 2013 Keck spectrum exhibits narrower Ca~II emission of 270 $\kmps$.  Fits to [Fe~II] $\lambda$7155 and [Ca~II] $\lambda\lambda$7291, 7324 show that no emission lines are detected in 2003, whereas the 2013 Keck spectrum exhibits strong, narrow emission consistent with the instrumental resolution (220 $\kmps$).
	
	We fit the spectrum blueward of 5000~\AA\ and find a best fit for a possible nonthermal power-law AGN contribution ($\lambda^{p}$) of $p=0.92 \pm 0.27$ in 2003 (SDSS) and $p=0.42 \pm 0.21$ in 2013 (Keck).

\subsection{Host Galaxy}

	SDSS1133 is located in Mrk 177, a nearby blue compact dwarf galaxy
with peculiar morphology \citep{Petrosian:2007:33}. The 5.8$\arcsec$ separation between SDSS1133 and the center of Mrk 177 corresponds to a projected physical distance of 0.81~kpc.  Given the observed magnitude of $g=15.69$ (SDSS), the galaxy absolute magnitude is $M_g =-16.6$, comparable to that of the Small Magellanic Cloud.  The center of the host galaxy was also observed spectroscopically by SDSS, resulting in a measurement of the oxygen abundance, 12 + log(O/H) = 8.58, somewhat oxygen-rich given its brightness, but still on the SDSS mass-metallicity relation from \citet{Tremonti:2004:898}.  In the host galaxy, Mrk 177, we measure the Ca~II NIR triplet absorption lines to be within $3 \pm 10~\kmps$ of the redshift derived from the [O~III] emission line.  

         We can also measure the star-formation rate (SFR) based on {\it GALEX} observations of Mrk 177 taken in 2004.  Because of the low spatial resolution of {\it GALEX}, we are unable to completely rule out the possibility of contamination by SDSS1133, but we measure a SFR of 0.05 M$_\odot$ yr$^{-1}$ based on \citet{Kennicutt:1998:541}.


\section{Discussion}
Table 3 summarizes the unusual properties of SDSS1133.  Given its long observed lifetime, luminosity, and highly variable behavior, SDSS1133 is peculiar among AGNs, as well as among SNe, tidal disruption flares, ULXs, and other stellar phenomena.  In the following sections, we discuss possible source scenarios for SDSS1133 in more detail. An LBV followed by a SN, or a recoiling SMBH, are the most likely scenarios, while a ULX, tidal flare, or tidally stripped AGN are disfavored by the observations.  Finally, we discuss the frequency of other events like SDSS1133.

\subsection{LBV Outburst, or LBV Outburst Followed by a SN}

	 SDSS1133 is a source with an optical absolute magnitude of $-13.5$ (which could either be constant or caught serendipitously erupting in 1950, 1994, and 1999) followed by a brief, luminous transient in 2001/2002 that reached $-16$ mag and slowly faded to $-13$ mag.  The value $-13.5$ mag exceeds the classical Eddington limit for a 100 $\Msun$ star.  The peak brightness of $-16$ is unprecedented for an LBV eruption, though there is still some debate about whether SN 2009ip and SN 1961V, both reaching nearly $-18$ mag in the optical, were extreme $\eta$~Carinae-like events or core-collapse SNe \citep[see also][]{Kochanek:2010, Smith:2011:773, VanDyk:2012:179}.  The closest LBV is $\eta$~Car, which had a ÒGreat EruptionÓ in the mid-19th century that was  $\sim -13.0$ mag from 1843 to 1855, with some uncertainty from the historical records and estimated extinction \citep{Smith:2011:2009}.  Broad-line emission is also found in blue compact dwarfs \citep{Izotov:2007:1297} because of stellar winds around young massive stars \citep{Izotov:2007:1297,Izotov:2012:1229} in regions of active star formation. However, stellar winds are ruled out for SDSS1133 since the H$\alpha$ luminosity of $0.7\times10^{40}$ erg s$^{-1}$ is too large.  A superbubble produced by multiple SNe is ruled out by the small spatial scale of the emission, $< 12$~pc.  As we discuss below, an LBV followed by a SN is a more likely scenario given the features of the recent optical spectra.

Late-time observations of interacting SNe provide critical information about the nature and mass-loss history of massive stars immediately before core collapse.  The redshifted portion of the H$\alpha$ and H$\beta$ lines has become suppressed over the last decade, and the reddest individual peak has diminished in luminosity between 2003 and 2013. This could be dust in the post-shock medium that is obscuring emission from ejecta components in the receding hemisphere of the explosion \citep[e.g., SN 2006jc;][]{Smith:2008:568}. The Type IIn SN 1998S \citep{Mauerhan:2012:2659} does exhibit [O~I], [O~II], and [O~III] emission 14 yr after outburst. Several SNe show ongoing broad emission as the fast SN ejecta cross the reverse shock behind the SN/CSM interface \citep[e.g., SNe 1980K, 1993J, 1970G, and 1957D;][]{Milisavljevic:2012:25}.  Finally, the appearance of late-time narrow [Fe~II] and [Ca~II] in the 2013 spectra has been found in other SNe \citep[e.g., SN 1992H;][]{Filippenko:1997:309}.

We can make a conservative order-of-magnitude estimate of the total bolometric luminosity of SDSS1133 over 12 yr based on the observed UV, optical, and NIR observations  (2200--24,000~\AA) to test whether it violates the total expected energy from a SN.  To do this, we assume that the spectral energy distribution (SED) of SDSS1133 follows the last UV observation in 2013, the 2014 MMT optical spectrum (3675--8850~\AA), and the $J$ and $K_p$ Keck measurements.  We interpolate between the unobserved regions in the range 2200--24,000~\AA.  We assume that the SED follows the dashed lines in the $g$-band light curve in Figure \ref{fig:bluephot} and obtain a total energy of $1.7 \times 10^{50}$ ergs.  This is probably a lower limit since we exclude the far-UV and X-ray emission, and the bolometric correction is larger at earlier times because of the greater amounts of UV and X-ray emission.  The largest amount of emission from a SN explosion thus far has been from SN 2003ma, with an integrated bolometric luminosity of $4 \times 10^{51}$ ergs.  While the integrated bolometric luminosity of SN 2003ma was a factor of $\sim 20$ higher than for SDSS1133, the peak emission of SN 2003ma was $M_R = -21.5$ mag, about 150 times brighter than SDSS1133. 

\begin{figure*} 
\centering
\includegraphics[width=8.1cm]{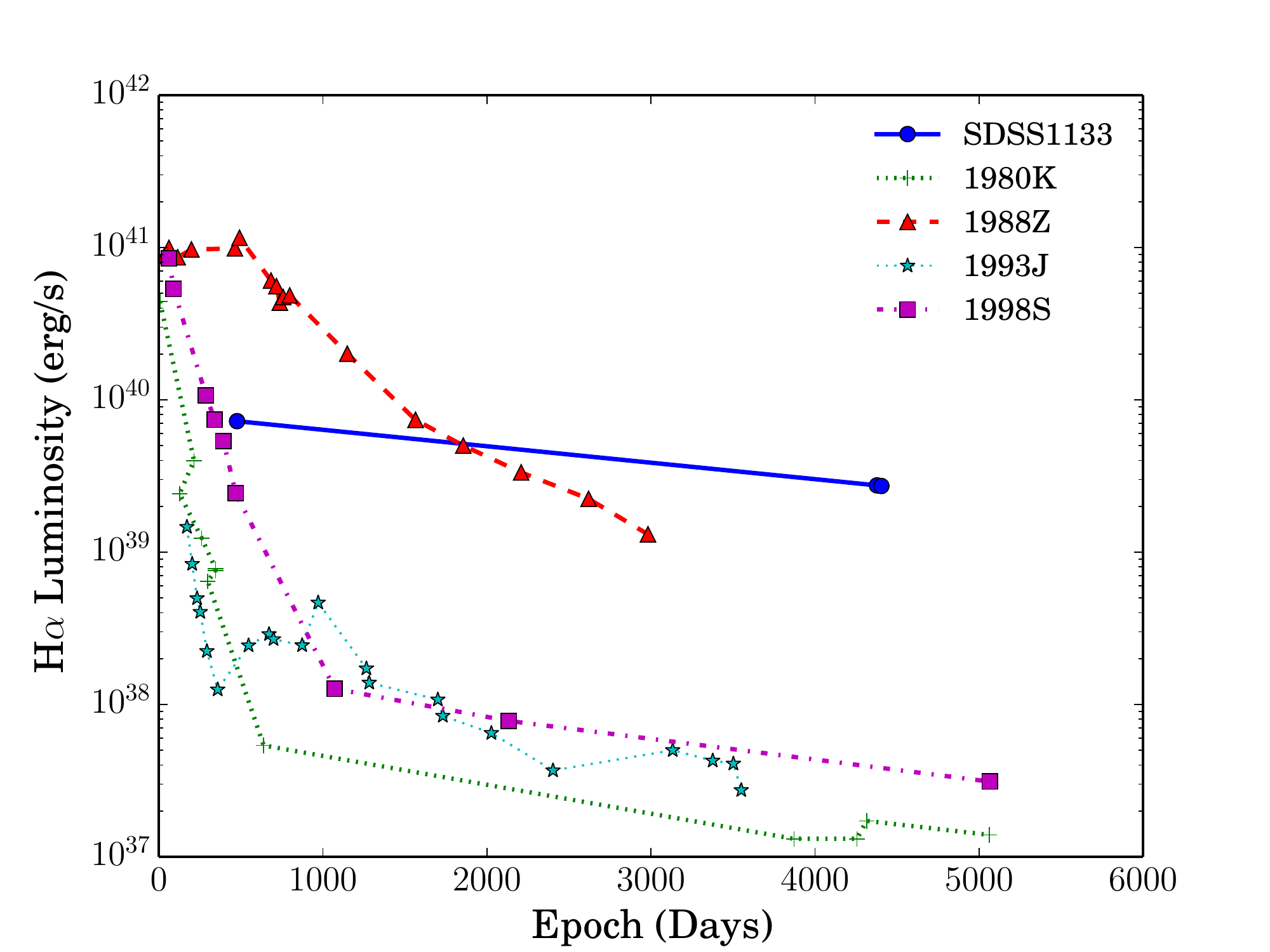}
\caption{H$\alpha$ emission from SDSS1133 compared to the most luminous late-time SNe from the literature  \citep{Aretxaga:1999:343,Mauerhan:2012:2659,Chandra:2009:388}.  SDSS1133 displays a very flat profile along with the brightest late-time H$\alpha$ ever seen, even including very luminous SNe such as SN 1988Z.}
\label{fig:SN_Halph}
\end{figure*}

The eruptive LBV and SN hypothesis has several issues that make SDSS1133 one of the most unusual LBV/SN candidates.  The 51 yr putative eruptive LBV phase before the SN (whose peak brightness was in 2001) is the longest observed before a SN explosion.  The luminous broad H$\alpha$ emission requires interaction with a very dense CSM linked to an extreme amount of mass loss by the LBV.  A plot of H$\alpha$ emission from SDSS1133 as a function of time after peak magnitude can be found in Figure~\ref{fig:SN_Halph}, as compared with the other most luminous SNe observed at late times.  In 2013, SDSS1133 shows little decrease in H$\alpha$ compared to other known SNe, and it has more luminous late-time H$\alpha$ emission than even extreme cases like SN 1988Z \citep{Aretxaga:1999:343}.  Additionally, UV emission in SNe typically drops by several magnitudes in tens of days, while the UV emission of SDSS1133 exhibits no change when observed between 2004 by {\it GALEX} and 2013 by {\it Swift}.  Strong UV emission is unexpected because SNe have line blanketing, and LBV eruptions usually form much dust.  A possible explanation is a cluster of O-type stars or an H~II region that is not detected in the NIR images.  However, the NIR AO Pa$\beta$ observations should directly trace the gas ionized by young, massive stars.  In this case, the flux would be overestimated, which would reduce the pre-SN flux and the late-time fading.  On the other hand, an H~II region seems unlikely given the $\sim 22$~pc resolution of AO in Pa$\beta$ and the change in narrow-line emission between 2003 and 2013.  Finally, additional monitoring is required to understand whether the light curve is consistent with a monotonic decrease expected from a SN.  If the variability is confirmed to be nonmonotonic, showing very significant rebrightening, scenarios for SDSS1133 with a LBV and SN origin will be severely constrained.

\subsection{An AGN}
\subsubsection{Recoiling Black Hole}

While the SMBH occupation fraction in dwarf galaxies is only beginning to be constrained \citep[e.g.,][]{Reines:2013:116}, any SMBH mergers that occur as a result of dwarf-dwarf mergers could produce a substantially offset recoiling AGN. This is because dwarf galaxies have low escape speeds, and thus even low-velocity recoil kicks may displace a SMBH from the galactic nucleus. If we assume that the potential of Mrk 177 is dominated by a dark-matter halo with an NFW profile \citep{Navarro:1996:563} for its spatial mass distribution of dark matter and a halo to stellar mass ratio of 30, then for the inferred stellar mass of $10^{8.55}~\Msun$, a reasonable guess for the central escape speed is $\sim 170$--200 km s$^{-1}$. The recent spectra have a broad H$\alpha$ component consistent with little or no velocity offset; thus, in the recoiling AGN scenario we can assume that either the recoil kick was directed largely in the transverse direction or that the AGN is on a bound orbit near turnaround. 

While high-velocity kicks ($> 1000 \kmps$) are expected to be quite rare, GW recoil velocities of 100--200 km s$^{-1}$ should be relatively common. \citet{Lousto:2012:84015} calculate that, even if the progenitor SMBH spins are mostly aligned prior to their merger (thus reducing the kick speed), 100--200 km s$^{-1}$ kicks should be produced in 21--28\% of SMBH mergers (assuming a cosmologically motivated distribution of SMBH mass ratios). These would correspond to bound or marginally unbound recoil kicks. An additional 12--13\% of such mergers should result in recoil velocities of 200--300 km s$^{-1}$, which would leave the recoiling AGN unbound from its host. 

The accretion rate onto a recoiling AGN should decrease monotonically with time as its accretion disk diffuses outward (unless the SMBH encounters a fresh fuel supply). Thus, even if the SMBH was accreting at the Eddington limit at the time of the kick, its accretion rate may be only a few percent of Eddington by the time it is observed as an offset AGN, several Myr later. This is consistent with the low AGN luminosity observed for SDSS1133. Note that typical offset AGN lifetimes are a few to a few tens of Myr \citep{Blecha:2011:2154}, consistent with the inferred travel time for a SMBH from the center of Mrk 177 to the current offset (0.8 kpc) of SDSS1133, and allowing for some deceleration. Furthermore, if the recoiling SMBH is on a bound orbit, it may be able to replenish its fuel supply on subsequent passages through the galaxy, thus extending its offset AGN lifetime \citep{Blecha:2011:2154}.

While the expected properties of recoiling AGNs are poorly understood, SDSS1133 does show many features similar to observed broad-line AGN and those expected for recoiling AGNs, including the following: (1) point-like emission for 63 yr that is on very small scales of $\lesssim 12$ pc,  (2) broad Fe~II emission is seen in almost all Type-1 AGN spectra,  (3) broad Ca~II NIR triplet emission is found in 1/3 of quasars \citep{Netzer:1990:57}, (4) a blue power-law continuum with no detected stellar absorption lines from the host-galaxy nucleus, (5) some narrow-line optical emission-line diagnostics consistent with an AGN, (6) a SMBH mass measurement that remains constant within the uncertainties over a decade, despite changes in the continuum,  (7) variability with dimming and rebrightening, (8) lack of significant color evolution, (9) relatively constant UV luminosity, and (10) H$\alpha$ luminosity consistent with a low-mass SMBH accreting at a typical rate of a few percent of $L_{\rm Edd}$.

Since the majority ($\sim 80$--100\%) of AGNs are variable \citep{MacLeod:2012:146}, and less luminous AGNs tend to be more variable \citep{Desjardins:2007:802}, one might expect SDSS1133 to show high variability.  However, the drop in brightness of SDSS1133 between the last SDSS image on 2002 April 1 and the SDSS spectra taken on 2003 March 9, over a period of about a year, is quite large at 2.2--2.9 mag.  While this sharp a change has been observed in a few AGNs \citep[e.g.,][]{Gaskell:2006:111}, the large variability and strong late-time UV emission are more consistent with tidal flares.

As an AGN, the rapid drop in luminosity between 2002 and 2003, the narrow [Fe~II] lines, and the presence of [Ca~II] $\lambda\lambda$7291, 7324 emission are difficult to understand.  The [Ca~II] lines are rarely seen in AGNs because dust grains are presumed to suppress the Ca~II emission through the substantial Ca depletion \citep{Ferland:1993:75}.  While found in some AGNs such as I~Zw~1 \citep{Phillips:1976:37}, the feature is rare among AGN spectra.

The total broad H$\alpha$ luminosity of $0.7\times10^{40} \ergps$ would make this a very low luminosity AGN, with a predicted bolometric luminosity of $\sim 10^{42} \ergps$ \citep{Greene:2007:131} and an Eddington ratio of 0.02.  Thus, this object is similar in its gross properties to many of the very low luminosity AGNs recently discovered, which have inferred BH masses in the $10^5$--$10^6~\Msun$ range \citep{Reines:2013:116}.  

Finally, in a variant of the recoiling AGN scenario, the extreme variability of SDSS1133 could arise from the tidal disruption of a star by the recoiling BH. This would result in a short-lived, luminous flare from the accretion of the stellar debris.  Tidal disruption events are bright in the X-rays, UV, and sometimes optical \citep[][]{Komossa:2002:436,Gezari:2006:L25}. For low-mass SMBHs such as SDSS1133, the flare is expected to be super-Eddington, with an estimated duration of $\sim 2$--5 yr \citep{Ulmer:1999:180}.  Stellar tidal disruptions by recoiling SMBHs may contribute a small but non-negligible fraction to the total rate of tidal disruptions by SMBHs \citep{Komossa:2008:L89,Stone:2012:1933}.

Because SDSS1133 is detected over 63 yr, with roughly constant luminosity and broad-line emission over the past decade, this scenario requires the SMBH to have a reservoir of gas persisting before and after the event, which produces continuous AGN emission. Since the 2003 spectrum does not show any distinct features expected from tidal disruption flares \citep[e.g.,][]{Strubbe:2011:168}, we conclude that any such flare must have faded in $< 2$ yr from observed peak brightness. This, combined with the small decline in brightness during the 104 days between the 2001 and 2002 observations, argues against the tidal disruption scenario.  More fundamentally, this scenario requires an additional rare event (a tidal disruption flare) in a highly unusual object, namely a recoiling BH in a dwarf galaxy.

\subsubsection{An Infalling Tidally Stripped Dwarf or ULX?}

It is theoretically possible that Mrk 177 is observed in an earlier stage of merging, such that SDSS1133 originated in a satellite galaxy falling into the host galaxy. In this case, the center of the host galaxy could contain its own, relatively quiescent SMBH.  The AO images in the $K_p$ and Pa$\beta$ bands both show an unresolved point source at the location of SDSS1133 down to spatial scales of 12~pc and 22~pc, respectively. This is consistent with a recoiling AGN that has left its host galaxy behind. If instead SDSS1133 is an infalling dwarf galaxy, then it must have been tidally stripped at radii $>12$~pc, a strong constraint. Numerical models suggest that repeated strong tidal encounters with a much more massive host can remove more than 99\% of the total initial mass of a satellite galaxy \citep[e.g.,][]{Penarrubia:2008:226}. However, the only galaxy near SDSS1133 is Mrk 177, a dwarf galaxy with insufficient mass to be responsible for such extreme tidal stripping of a fellow dwarf companion. 

When an SMBH is present, its sphere of influence sets an upper limit on the amount of stripping; even strong tidal encounters will be ineffective at removing mass where the SMBH dominates the gravitational potential.  Making the first-order assumption that the dwarf galaxy followed the observed SMBH-bulge scaling relations prior to stripping, our SMBH mass estimate of $10^{6.0}~\Msun$  for SDSS1133 implies a stellar velocity dispersion of 60 km s$^{-1}$ \citep{Tremaine:2002:740}. Using the standard definition for the size of the SMBH sphere of influence, $r_{\rm infl} = G M_{\rm BH} / \sigma_*^2$, we find $r_{\rm infl} \approx 1$ pc. This implies that the SMBH sphere of influence is likely to be unresolved, such that a tidal dwarf remnant stripped to within a few times $r_{\rm infl}$ cannot be ruled out. Note that extrapolation from the SMBH-bulge scaling relations \citep{Haring:2004:L89} implies a progenitor galaxy for SDSS1133 \emph{larger} than Mrk 177. However, the value of $\sigma_*$ inferred from \citet{Tremaine:2002:740} is likely an upper limit. Tidal stripping and stellar feedback can decrease the velocity dispersion  \citep[e.g.,][]{Penarrubia:2009:222,Navarro:1996:L72, Zolotov:2012:71}, and the progenitor may also have been somewhat less massive. All of these would yield a larger radius of influence. Specifically, a factor of 4 decrease in velocity dispersion from the above estimates would imply that the SMBH sphere of influence is at least marginally resolvable ($r_{\rm infl} \approx 7$--19 pc).  Therefore, the tidal stripping scenario is very strongly constrained by the need for both an unidentified massive galaxy to strip the SDSS1133 progenitor and for a compact, unresolvable SMBH sphere of influence.

In contrast, a recoiling SMBH ejected from a spherical stellar distribution will only retain stars within a radius $r \approx (\sigma_* / v_{\rm k})^2 \, r_{\rm infl}$. For a Bahcall-Wolf profile normalized such that $M(r<r_{\rm infl}) = 2 M_{\rm BH}$ \citep{Bahcall:1976:214}, a kick of 200 km s$^{-1}$ would only allow the inner 0.003 pc of the stellar cusp, or a few thousand $\Msun$, to remain bound to the SMBH. Thus, in the recoil scenario the stellar cusp would not be resolved under any circumstances. 

Interpreting SDSS1133 as a ULX, it is either accreting ambient gas (i.e., indistinguishable from an AGN discussed in previous scenarios) or accreting from a stellar companion (which is untenable for our large inferred BH mass $10^6~\Msun$).  The most extreme ULX is HLX-1 (ESO 243-49), with $L_{\rm X} = 2\times10^{42} \ergps$ and a BH mass of (3--300) $\times 10^3~\Msun$ \citep{Farrell:2009:73}, making it the most convincing example of an IMBH.  This BH mass is significantly lower than that inferred for SDSS1133. HLX-1 is offset from its host-galaxy center by $\sim 3$ kpc (in projection), and may reside in a tidally stripped dwarf galaxy that is merging with the host galaxy \citep[e.g.,][]{Mapelli:2012:1309}. In principle, the AGN in SDSS1133 could have a similar origin, such that its host galaxy was tidally stripped long ago, but the point-like AO observations and lack of nearby massive galaxies place strong constraints against this.   If SDSS1133 were a ULX, it would be the first ever to be observed with broad Balmer lines \citep{Roberts:2011:398}.  Given the inferred BH mass of SDSS1133, as a ULX it is extremely X-ray weak, with a 2--10 keV to $V$-band flux ratio of $\sim 100$; this is below that of most ULXs, though some in their low states such as M101-ULX1 and M81-ULX1 have observed ratios similar to this \citep{Tao:2012:85}.  

\subsection{Frequency of Objects like SDSS1133}

 A survey of all 3579 low-redshift ($z<0.3$) broad emission-line objects ($> 1000~\kmps$ FWHM; \citealt{Stern:2012:600}) found only two objects offset from the host-galaxy nucleus: SDSS1133 along with a very close dual AGN Mrk 739 \citep{Koss:2011:L42}.  The low frequency of detections of spatially offset broad-line sources found in the SDSS quasar survey (1/3579, $<0.1$\%) does not exclude the possibility of a significant population.  The small physical separation of 800 pc found for SDSS1133 can only be resolved at very low redshift with the imaging of the SDSS (at $z<0.03$, 1.3$\arcsec \approx 800$ pc).   The broad-line luminosity of SDSS1133 is among the weakest dozen of the SDSS sample, and at the current reduced brightness of SDSS1133 it would be too faint for the survey \citep{Richards:2004:46}.  Future space-based surveys such as $Euclid$ and $WFIRST$ will obtain high-resolution images with precision astrometry of large areas of the sky, enabling them to probe the nuclear regions of nearby galaxies for offset point sources. Both have IR grisms as part of their wide-field surveys to cover broad AGN lines in the NIR such as the Paschen lines.

\section{Conclusion}

SDSS1133 has many observed properties consistent with a recoiling AGN, but it also has some properties that favor interpretation of this source as an eruptive luminous blue variable star outbursting for decades followed by a Type IIn SN.  SDSS1133 would be the longest LBV eruption ever observed and have a much greater late-time H$\alpha$ luminosity than even extreme events like SN 1988Z.  SDSS1133 has recently undergone a 1 mag rebrightening in PS1 images, suggesting that the coming years will be critical to understand the true nature of this source.  Future high spatial resolution and high sensitivity UV, X-ray, and radio observations of SDSS1133 are critical for constraining the source nature.  In the UV, the measurement of strong N~V $\lambda$1240 and (especially) broad C~IV $\lambda$1550 emission would decisively favor the AGN interpretation.  Additionally, high-resolution UV images could determine whether the strong UV emission and prominent narrow-line emission in [O~I], [O~II], and [O~III] is produced by a compact H~II region or a young star cluster not seen in the NIR AO image.  High-sensitivity X-ray observations could differentiate between softer thermal X-ray emission from a SN shock and an AGN power law.
	
\section*{Acknowledgments}

We are grateful to Jessica Lu and Aaron Barth for useful discussion and suggestions.  We thank Neil Gehrels and the {\it Swift} team for approving and executing a ToO observation.  M.K. and K.S. acknowledge support from Swiss National Science Foundation (NSF) grant PP00P2 138979/1.  M.K. also acknowledges support for this work provided by the National Aeronautics and Space Administration (NASA) through Chandra Award Number AR3-14010X issued by the Chandra X-ray Observatory Center, which is operated by the Smithsonian Astrophysical Observatory for and on behalf of NASA under contract NAS8-03060.  Support for L.B. was provided by NASA through the Einstein Fellowship Program, grant PF2-130093.  A.V.F. received generous financial assistance from Gary and Cynthia Bengier, the Christopher R. Redlich Fund, the Richard and Rhoda Goldman Fund, the TABASGO Foundation, and NSF grant AST-1211916.  The work of D.S. was carried out at Jet Propulsion Laboratory, California Institute of Technology, under a contract with NASA.  

Some of the data presented herein were obtained at the W. M. Keck Observatory, which is operated as a scientific partnership among the California Institute of Technology, the University of California, and NASA; the Observatory was made possible by the generous financial support of the W. M. Keck Foundation. Data reported here were obtained in part at the MMT Observatory, a joint facility of the University of Arizona and the Smithsonian Institution.  Funding for the SDSS and SDSS-II has been provided by the Alfred P. Sloan Foundation, the Participating Institutions, the NSF, the U.S. Department of Energy, NASA, the Japanese Monbukagakusho, the Max Planck Society, and the Higher Education Funding Council for England. The SDSS website is http://www.sdss.org/.  The PS1 data have been made possible through contributions of the Institute for Astronomy, the University of Hawaii, the Pan-STARRS1 Project Office, the Max-Planck Society and its participating institutes, the Max Planck Institute for Astronomy (Heidelberg) and the Max Planck Institute for Extraterrestrial Physics (Garching), The Johns Hopkins University, Durham University, the University of Edinburgh, Queen's University Belfast, the Harvard-Smithsonian Center for Astrophysics, the Las Cumbres Observatory Global Telescope Network Incorporated, the National Central University of Taiwan, the Space Telescope Science Institute, NASA under grant NNX08AR22G issued through the Planetary Science Division of the NASA Science Mission Directorate, the NSF under grant AST-1238877, the University of Maryland, and Eotvos Lorand University (ELTE). 

This research made use of the NASA/IPAC Extragalactic Database (NED) which is operated by the Jet Propulsion Laboratory, California Institute of Technology, under contract with NASA. It also employed Astropy, a community-developed core Python package for Astronomy (Astropy Collaboration, 2013). Moreover, it used APLpy, an open-source plotting package for Python hosted at http://aplpy.github.com.


\footnotesize{

}


\begin{table*}
\centering
 \begin{minipage}{140mm}
\label{table:phot}
\caption{Photometry}
\begin{tabular}{cccccc}
\hline \hline
Filter \footnote{Specific filter used for image.  The 103aO, IIIaJ, and IIIaF designations refer to plates from the DSS.}& Mag \footnote{All magnitudes are listed as AB mag.} & Mag err \footnote{Uncertainty from photometric zero point and uncertainty in measuring photometry with GALFIT.  For synthetic photometry from spectra, the uncertainty includes the conversion from aperture to PSF magnitudes.} & Date & Telescope \footnote{MMTspec and SDSSspec indicate synthetic photometry; all of the other observations were from imaging.}  \footnote{Number of observations taken at separate times within 1 week of the first observation.  For PS1 several observations are often taken within a week; we have measured the weighted mean of these separate observations.}  \\
\hline
103aO & 18.6 & 0.7 & 1950-03-20 & DSS1 & 1 \\
IIIaJ & 18.4 & 0.40 & 1994-04-14 & DSS2 & 1 \\
IIIaF & 18.8 & 0.50 & 1999-04-25 & DSS2 & 1 \\
u & 16.28 & 0.08 & 2001-12-18 & SDSS & 1 \\
u & 16.75 & 0.08 & 2002-04-01 & SDSS & 1 \\
g & 16.41 & 0.04 & 2001-12-18 & SDSS & 1 \\
g & 16.40 & 0.04 & 2002-04-01 & SDSS & 1 \\
g & 18.7 & 0.18 & 2003-03-09 & SDSSspec & 1 \\
i & 16.48 & 0.04 & 2001-12-18 & SDSS & 1 \\
i & 16.40 & 0.04 & 2002-04-01 & SDSS & 1 \\
i & 18.65 & 0.05 & 2003-03-09 & SDSSspec & 1 \\
r & 16.26 & 0.04 & 2001-12-18 & SDSS & 1 \\
r & 16.31 & 0.04 & 2002-04-01 & SDSS & 1 \\
r & 18.28 & 0.06 & 2003-03-09 & SDSSspec & 1 \\
z & 16.45 & 0.04 & 2001-12-18 & SDSS & 1 \\
z & 16.34 & 0.04 & 2002-04-01 & SDSS & 1 \\
z & 18.49 & 0.08 & 2003-03-09 & SDSSspec & 1 \\
NUV & 21.62 & 0.40 & 2004-03-06 & GALEX & 1 \\
g & 19.4 & 0.0 & 2005-01-01 & Beijing & 1 \\
g & 19.58 & 0.01 & 2010-03-17 & PS1 & 4 \\
g & 19.68 & 0.05 & 2011-03-12 & PS1 & 8 \\
g & 20.18 & 0.01 & 2012-02-22 & PS1 & 2 \\
g & 19.38 & 0.07 & 2013-01-16 & PS1 & 8 \\
r & 19.02 & 0.03 & 2010-03-13 & PS1 & 4 \\
r & 19.37 & 0.01 & 2011-03-12 & PS1 & 4 \\
r & 19.91 & 0.04 & 2012-02-22 & PS1 & 4 \\
r & 19.31 & 0.06 & 2012-12-27 & PS1 & 4 \\
r & 19.49 & 0.02 & 2013-02-09 & PS1 & 8 \\
r & 19.02 & 0.03 & 2014-03-24 & PS1 & 4 \\
i & 19.48 & 0.03 & 2010-03-03 & PS1 & 4 \\
i & 19.5 & 0.03 & 2011-03-14 & PS1 & 4 \\
i & 20.12 & 0.03 & 2012-02-11 & PS1 & 4 \\
i & 19.62 & 0.04 & 2013-01-27 & PS1 & 4 \\
i & 19.48 & 0.03 & 2014-03-16 & PS1 & 4 \\
z & 19.43 & 0.07 & 2010-02-27 & PS1 & 8 \\
z & 19.72 & 0.03 & 2010-05-18 & PS1 & 8 \\
z & 19.57 & 0.01 & 2010-12-25 & PS1 & 4 \\
z & 19.73 & 0.10 & 2011-01-22 & PS1 & 4 \\
z & 19.82 & 0.01 & 2011-05-11 & PS1 & 4 \\
z & 20.11 & 0.12 & 2012-02-09 & PS1 & 4 \\
z & 19.82 & 0.02 & 2012-04-10 & PS1 & 4 \\
z & 19.48 & 0.02 & 2013-01-03 & PS1 & 4 \\
z & 19.32 & 0.05 & 2014-01-20 & PS1 & 1 \\
J & 19.02 & 0.18 & 2013-06-16 & NIRC2 & 1 \\
Kp & 18.92 & 0.16 & 2013-06-16 & NIRC2 & 1 \\
uvw1 & 21.41 & 0.30 & 2013-08-27 & SWIFT & 1 \\
g & 19.19 & 0.20 & 2014-01-04 & MMTspec & 1 \\
\hline
\end{tabular}
\end{minipage}
\end{table*}

\begin{table*}
\label{table:em_line}
 \centering
 \begin{minipage}{140mm}
  \caption{Emission-Line Properties}
  \begin{tabular}{@{}lllllll@{}}
  \hline

Name\footnote{Name of spectrum.  Mrk 177 and SDSS1133 are from the SDSS.  Mrk 177 IFU-offset is taken using the SNIFS IFU at the same radial offset from the nucleus of Mrk 177 as SDSS1133 (but in the opposite direction) and with the same aperture size.  SDSS1133-IFU is taken of SDSS1133 with Mrk 177 IFU-offset subtracted, leaving only the broad lines to fit.} &Line &Offset \footnote{Offset from the measured [O~III] line in Mrk 177.} &FWHM \footnote{All of the narrow lines measured were consistent with the instrumental resolution.} &Luminosity \footnote{Luminosity assuming a distance of 28.9 Mpc.} &Date\\
 & &($\kmps$) &($\kmps$) &($10^{36~\ergps}$) &\\

\hline
Narrow Lines\\
\hline
SDSS1133& H$\beta $                  		& 26$\pm$9 & 150 & 180$\pm$8&2003 \\
SDSS1133& H$\beta $                  		& 31$\pm$15 & 220 & 94$\pm$6&2013 \\

SDSS1133& [O III] $\lambda$5007 		& 26$\pm$9 & 150 & 391$\pm$3&2003 \\
SDSS1133& [O III] $\lambda$5007 		& 31$\pm$15 & 220 & 170$\pm$3&2013 \\

SDSS1133& [O I] $\lambda$6300 		& 26$\pm$9 & 150 & 16$\pm$6&2003 \\
SDSS1133& [O I] $\lambda$6300 		& 31$\pm$9 & 220 & 22$\pm$2&2013 \\

SDSS1133& H$\alpha$ 				& 26$\pm$9 & 150 & 211$\pm$7&2003 \\
SDSS1133& H$\alpha$ 				& 31$\pm$9 & 220 & 335$\pm$21&2013 \\

SDSS1133& [N II] $\lambda$6583 		& 26$\pm$9 & 150 & 29$\pm$2&2003 \\
SDSS1133& [N II] $\lambda$6583 		& 31$\pm$9 & 220 & 15$\pm$8&2013 \\

SDSS1133& [S II] $\lambda$6716		& 26$\pm$9 & 150 & 51$\pm$6&2003 \\
SDSS1133& [S II] $\lambda$6716		& 31$\pm$9 & 220 & 27$\pm$3&2013 \\

SDSS1133& [S II] $\lambda$6731		& 26$\pm$9 & 150 & 34$\pm$6&2003 \\
SDSS1133& [S II] $\lambda$6731		& 31$\pm$9 & 220 & 19$\pm$2&2013 \\
SDSS1133& [Fe II] $\lambda$7155		& 19$\pm$7 & 220 & 15$\pm$2 &2013\\

SDSS1133& [Ca II] $\lambda$7291		& 14$\pm$4 & 220 & 46$\pm$3 &2013\\
SDSS1133& [Ca II] $\lambda$7324		& 8$\pm$4 & 220 & 38$\pm$3 &2013\\

\hline
Broad Lines\\
\hline
SDSS1133& H$\beta$ BLR             		& 252$\pm$43 & 1623$\pm$144 & 554$\pm$66&2003 \\
SDSS1133& H$\beta$ BLR             		& 252$\pm$43 & 1623$\pm$144 & 441$\pm$42&2013 \\

SDSS1133& H$\alpha$ BLR 			& 330$\pm$8 & 1254$\pm$26 & 2335$\pm$89&2003 \\
SDSS1133& H$\alpha$ VBLR 			& 876$\pm$28 & 4397$\pm$72 & 4703$\pm$201&2003 \\

SDSS1133& H$\alpha$ BLR 				& 285$\pm$5 & 2310$\pm$13 & 1555$\pm$17&2013 \\
SDSS1133& H$\alpha$ VBLR 				& 1888$\pm$38 & 6567$\pm$67 & 862$\pm$26&2013 \\
SDSS1133& H$\alpha$ BLR 				& 38$\pm$5 & 879$\pm$35 & 578$\pm$89 &2014 \\
SDSS1133& H$\alpha$ VBLR 				& 72$\pm$8 & 3281$\pm$532 & 1698$\pm$342 &2014 \\

SDSS1133& O I $\lambda$8446		& 328$\pm$36 & 647$\pm$42 & 72$\pm$16 &2003\\
SDSS1133& O I $\lambda$8446		& 41$\pm$5 & 262$\pm$3 & 11$\pm$2 &2013\\
SDSS1133& Ca II $\lambda$8498		& 176$\pm$22 & 643$\pm$43 & 72$\pm$15&2003 \\
SDSS1133& Ca II $\lambda$8498		& 38$\pm$6 & 264$\pm$10 & 27$\pm$3 &2013\\
SDSS1133& Ca II $\lambda$8452		& 176$\pm$22 & 640$\pm$43 & 79$\pm$10&2003 \\
SDSS1133& Ca II $\lambda$8452		& 38$\pm$4 & 265$\pm$10 & 28$\pm$3 &2013\\
SDSS1133& Ca II $\lambda$8662		& 173$\pm$22 & 631$\pm$43 & 53$\pm$11&2003 \\
SDSS1133& Ca II $\lambda$8662		& 37$\pm$4 & 265$\pm$3 & 23$\pm$3 &2013\\

\hline
Host Galaxy\\
\hline

Mrk 177& H$\beta $                  		& 0 & 150 & 115$\pm$9&2003 \\
Mrk 177-IFU offset& H$\beta $                  		& -18$\pm$6 & 360 & 103$\pm$35&2013 \\
Mrk 177& [O III] $\lambda$5007 		& 0 & 150 & 188$\pm$4&2003 \\
Mrk 177-IFU offset& [O III] $\lambda$5007 		& -18$\pm$6 & 360 & 69$\pm$25&2013 \\
Mrk 177& [O I] $\lambda$6300 		& 0 & 150 & 26$\pm$5&2003 \\
Mrk 177-IFU offset& [O I] $\lambda$6300 		& -18$\pm$6 & 360 & 6$\pm$3&2013 \\
Mrk 177& H$\alpha$ 				& 0 & 150 & 332$\pm$5&2003 \\
Mrk 177-IFU offset& H$\alpha$ 				& -18$\pm$6 & 360 & 127$\pm$17&2013 \\
Mrk 177& [N II] $\lambda$6583 		& 0 & 150 & 86$\pm$5&2003 \\
Mrk 177-IFU offset& [N II] $\lambda$6583 		& -18$\pm$6 & 360 & 25$\pm$8 &2013\\
Mrk 177& [S II] $\lambda$6716		& 0 & 150 & 86$\pm$5&2003\\
Mrk 177-IFU offset& [S II] $\lambda$6716		& -18$\pm$6 & 360 & 27$\pm$11 &2013\\
Mrk 177& [S II] $\lambda$6731		& 0 & 150 & 63$\pm$5&2003 \\
Mrk 177-IFU offset& [S II] $\lambda$6731		& -18$\pm$6 & 360 & 22$\pm$11 &2013\\
\hline
\end{tabular}
\end{minipage}
\end{table*}

\begin{table*}
\label{table:source_prop}
 \centering
 \begin{minipage}{130mm}
  \caption{Source Properties}
  \begin{tabular}{@{}lccc@{}}
  \hline

Property\footnote{R indicates rarely, Y indicates common, and blank indicates never \\ observed among this type of source.} &AGN &LBV Outburst &SNe\\
\hline

Bright ($>-13$ mag) optical emission for 51 yr&Y&Y&\\
Optical drop of 2.2 mag in 1 year&R&Y&Y\\
Peak optical mag of $-16$ &Y&&Y\\
 $< 20$ pc point-source emission&Y&Y&Y\\
Constant UV emission for a decade&Y&&\\
Constant $g-i$ color for a decade&Y&&\\
Changing broad-line width&Y&&Y\\
Broad H$\alpha$ ($> 1000~\kmps$) for a decade&Y&&R\\
Broad-line Balmer decrement change&R&&Y\\
Narrow [O~III] emission&Y&&R\\
Late-time narrow [Fe~II] emission&R&&R\\
Late-time narrow [Ca~II] emission&R&&R\\
P-Cygni blueshifted absorption&R&&Y\\
$i-K_p$ color = 2.5 mag &Y&&R\\
0.3--10 keV/H$\alpha$ ratio $\approx$ 1&R&&Y\\
\hline
\end{tabular}
\end{minipage}
\end{table*}

\end{document}